\documentclass[aps,pra,twocolumn,superscriptaddress,nofootinbib]{revtex4-1}
\usepackage{braket}
\usepackage{graphicx}
\usepackage{amsmath}    % need for subequations
\usepackage{amsfonts}
\usepackage{subfigure}  % use for side-by-side figures
\usepackage{epstopdf}
\usepackage{multirow}

\begin{document}

\title{Ground-state cooling enabled by \\ critical coupling and dark entangled states}
\author{Cristian L. Cortes}
\affiliation{Center for Nanoscale Materials,\\ Argonne National Laboratory, Lemont, Illinois 60439, USA}
\author{Matthew Otten}
\affiliation{Center for Nanoscale Materials,\\ Argonne National Laboratory, Lemont, Illinois 60439, USA}
\author{Stephen K. Gray}
\affiliation{Center for Nanoscale Materials,\\ Argonne National Laboratory, Lemont, Illinois 60439, USA}

\begin{abstract}
We analyze the cooling of a mechanical resonator coupled to an ensemble of interacting two-level systems via an open quantum systems approach. Using an exact analytical result, we find optimal cooling occurs when the phonon mode is critically coupled ($\gamma \sim g$) to the two-level system ensemble. Typical systems operate in sub-optimal cooling regimes due to the intrinsic parameter mismatch ($\gamma \gg g$) between the dissipative decay rate $\gamma$ and the coupling factor $g$. To overcome this obstacle, we show that carefully engineering the coupling parameters through the strain profile of the mechanical resonator allows phonon cooling to proceed through the dark (subradiant) entangled states of an \emph{interacting} ensemble, thereby resulting in optimal phonon cooling. %Our result points towards the control of the dissipative decay rate as an important design principle for resonant phonon cooling.
Our results provide a new avenue for ground-state cooling and should be accessible for experimental demonstrations.

\end{abstract}

\maketitle

% \blfootnote{The submitted manuscript has been created by UChicago Argonne, LLC, Operator of Argonne
% National Laboratory ("Argonne"). Argonne, a U.S. Department of Energy Offce of Science laboratory, is operated under Contract No. DE-AC02-06CH11357. The U.S. Government retains for itself, and others acting on its behalf, a paid-up nonexclusive, irrevocable worldwide license in said article to reproduce, prepare derivative works, distribute copies to the public, and perform publicly and display publicly, by or on behalf of the Government.}

\section{Introduction}

A nanomechanical resonator reaches the quantum regime when the phonon occupation number falls below one, $\braket{a^\dagger a} \lesssim 1$. Fundamentally, this indicates the mechanical mode has at most one quantum of energy, making quantum coherence and entanglement observable while enabling the study of quantum nanomechanics with macroscopic objects \cite{jost2009entangled,o2010quantum,aspelmeyer2014cavity}. A resonator with near-zero thermal noise will have better performance characteristics in nanoscale sensing, quantum memories, and quantum information processing applications \cite{lahaye2004approaching,cleland2004superconducting}.

One obvious way of entering the quantum regime is through cryogenic cooling. In practice, passive cooling systems, such as dilution refrigerators, have a low temperature limit of around $2-10$ mK. While these temperatures allow for ground-state cooling of high-frequency resonators ($\omega_m/2\pi \gtrsim 1$ GHz), additional cooling techniques are still required for systems with lower frequencies. To this end, the most successful cooling techniques have been based on the optomechanical effect. The mechanical resonator is parametrically coupled to a driven optical cavity \cite{wilson2004laser,martin2004ground,PhysRevLett.99.093901,PhysRevLett.99.093902,chan2011laser} and,  under stringent conditions, careful tuning of the drive frequency results in resonator cooling in a form  analogous to the laser cooling of trapped atoms or ions \cite{cirac1992laser,diedrich1989laser}. 

Alternative cooling techniques have also been proposed based on direct (non-parametric) coupling between the mechanical resonator and a dissipative two-level system (TLS) \cite{kepesidis2013phonon,macquarrie2013mechanical,macquarrie2017cooling}. The two-level system may represent color defects in diamond or quantum dots coupled to the mechanical resonator using either spin-strain or orbital-strain interactions. In this approach, a complete description of the cooling process requires the atom to be modeled as a three-level system (see Figure 1-a and Appendix A for details) \cite{kepesidis2013phonon}. Here, an external laser pumps the atom into the ground state of a magnetically-forbidden two-level subspace. The coupling between the two-level subspace and the mechanical resonator induces a transition from the ground-state $\ket{g}$ to the excited state $\ket{e}$ due to phonon absorption. The atom then  releases the excess energy in the form of a high-energy photon and effectively cools the resonator by a single quantum. After many cycles, and under specific conditions which we outline in this manuscript, the mechanical resonator eventually reaches its ground state. The main advantage of this approach is the potential for miniaturization and room-temperature operation required for long-term technological applications. The two-level system also brings additional functionalities that may be used once the system is cooled. For example, the TLS can serve as a sensor, or be used to prepare non-classical phonon states \cite{meekhof1996generation,cirac1995quantum,monroe1995demonstration,armour2002entanglement}. Due to the intrinsically small spin-strain interaction in real systems, there have not been experimental demonstrations of phonon cooling using this approach. 

In this manuscript, we develop a quantum theory of phonon cooling using an \emph{interacting} two-level system ensemble, depicted in Figure 1-b. We reveal critical coupling as an important and universal condition for optimizing phonon cooling. We also propose using the strain profile of  the mechanical resonator to \emph{selectively} couple to the dark (subradiant) entangled states of an interacting two-level ensemble. This approach allows the critical coupling condition to be fulfilled and overcomes the major obstacles that have prevented the realization of phonon cooling using embedded solid-state defects. 

In our theoretical framework, we utilize the Jaynes-Cummings Hamiltonian which describes a wide variety of two-level systems ranging from quantum dots, superconducting qubits, and NV centers in diamond. While we have focused our analysis on the cooling of a single mechanical mode, our results are also applicable to many other low-frequency bosonic systems such as magnons, phonon-polaritons, and low-frequency plasmons. The paper is structured in order of increasing complexity. In Sec. 1, we investigate the parameter regimes required for optimal cooling for a single dissipative two-level system. The main result is the critical coupling condition required for optimal cooling. In Sec. 2, we generalize the theory to $N$ non-interacting two-level systems. In Sec. 3, we develop the theory for $N$ interacting two-level systems where we demonstrate that subradiant eigenstates enhance the phonon cooling performance. Finally, we discuss the role of entanglement, practical implementations, as well as the effects of noise and disorder. 

\begin{figure}
  \centering
  \includegraphics[width=9cm]{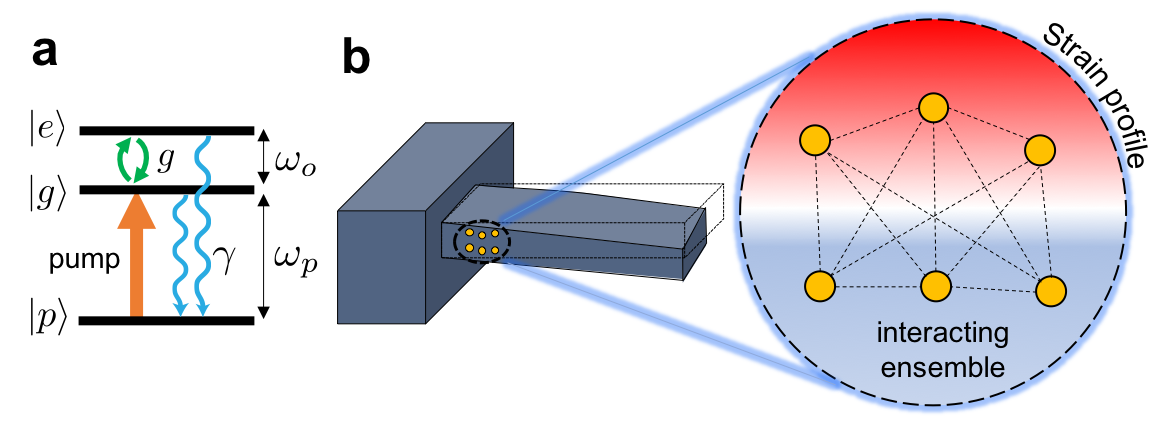}
\caption{(a) Model for ground-state cooling using phonon-assisted optical transitions in a three-level system. An external laser pumps the system into the ground state $\ket{g}$ of a magnetically-forbidden two-level subspace ($\omega_o\ll\omega_p$) with decay rate $\gamma$. The two-level subspace couples directly to the mechanical resonator with coupling strength $g$. Critical coupling ($g \sim \gamma$) ensures efficient cooling due to optimal phonon absorption from the ground state $\ket{g}$ to the excited state $\ket{e}$. (b) Schematic of resonant phonon cooling using  odd-parity strain profile. This configuration couples directly to the dark (subradiant) eigenstates of a two-level system ensemble with all-to-all interactions resulting in optimal cooling. 
}
\label{fig:test}
\end{figure}

% \begin{figure*}
% \centering
% \begin{subfigure}
%   \centering
%   \includegraphics[width=.45\linewidth]{NewFigure1aa.png}
%   \label{fig:sub1}
% \end{subfigure}%
% \begin{subfigure}
%   \centering
%   \includegraphics[width=.45\linewidth]{NewFigure1b.png}
%   \label{fig:sub2}
% \end{subfigure}
% \caption{(a) Schematic of resonant phonon cooling. (b) Phonon cooling mechanism through two-atom Jaynes-Cummings ladder. The critical coupling condition effectively results in one-way state transfer down the ladder from $\ket{n}$ to $\ket{n-1}$ phonon Fock states (solid lines). Dashed lines correspond to cooling mechanism giving rise to shoulder in Figure 3. }
% \label{fig:test}
% \end{figure*}

\section{Cooling with a single atom}

We first consider the resonant cooling of a phonon mode coupled to a  single two-level system. We emphasize that the effective two-level system description arises from the three-level model with incoherent pumping. Details are included in appendix A. The two-level system system is described by the Jaynes-Cummings Hamiltonian ($\hbar = 1$)
\begin{equation}
	H = \frac{\omega_o}{2} \sigma_z + \omega_m a^\dagger a + g(a^\dagger \sigma + a \sigma^\dagger)
\end{equation}
where $\omega_o$ and $\omega_m$ are the TLS transition frequency and the phonon mode frequency respectively. We also define $\sigma^\dagger = \ket{e}\bra{g}$ and $\sigma = \ket{g}\bra{e}$ as the Pauli raising and lowering operators of the TLS, along with $\sigma_z = \ket{e}\!\bra{e}- \ket{g}\!\bra{g}$. The mechanical mode is described by creation and annihilation operators $a^\dagger$ and $a$, and interacts with the TLS with coupling strength $g$. The total open quantum system is described by the density operator $\rho$ obeying the Lindblad master equation,
\begin{align}
	\dot\rho &= -i[H,\rho] + \gamma \mathcal{D}[\sigma]\rho + \gamma_\phi \mathcal{D}[\sigma_z]\rho \ \nonumber \\
	&+ \kappa(n_{th}+1)\mathcal{D}[a]\rho + \kappa n_{th}\mathcal{D}[a^\dagger]\rho
\end{align}
where $\mathcal{D}[o]\rho = o\rho o^\dagger - \frac{1}{2}(o^\dagger o \rho + \rho o^\dagger o)$ is the dissipative Lindblad superoperator acting the density matrix $\rho$. We assume the mechanical resonator is in contact with a thermal bath with temperature $T$ and occupation number $n_{th} = (e^{\hbar\omega_m/kT}-1)^{-1}$, where $\kappa$ denotes the mechanical damping rate. We have also included a dephasing rate $\gamma_\phi$ and spontaneous emission rate $\gamma$ for the two-level system. The two-level system does not include Lindblad terms for an electrodynamic thermal bath as the two-level subspace is assumed to be optically forbidden.

\begin{figure*}
	 \centering
  \begin{minipage}[b]{0.495\textwidth}
    \includegraphics[width=\textwidth]{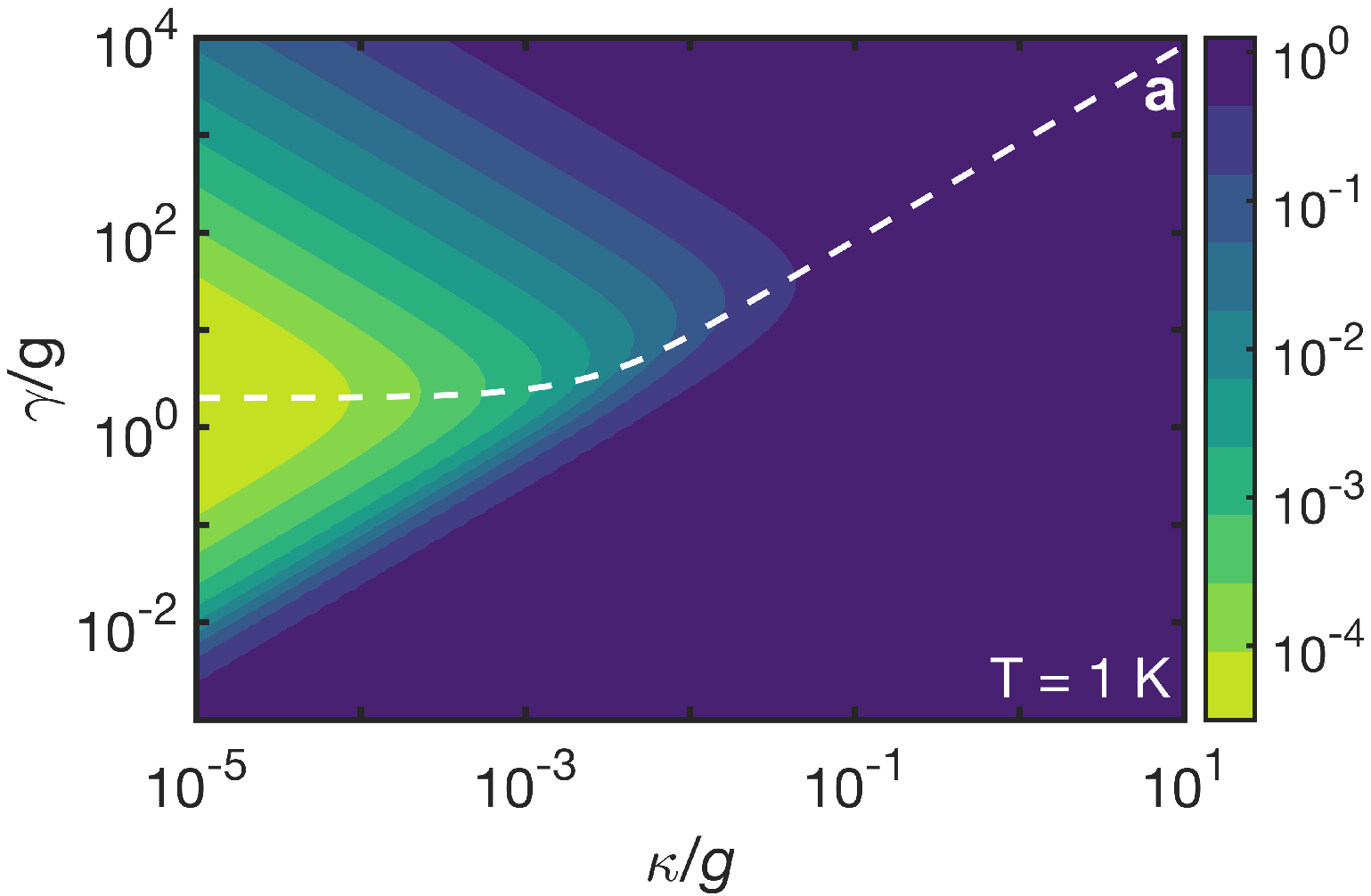}
  \end{minipage}
  \hfill
  \begin{minipage}[b]{0.49\textwidth}
    \includegraphics[width=\textwidth]{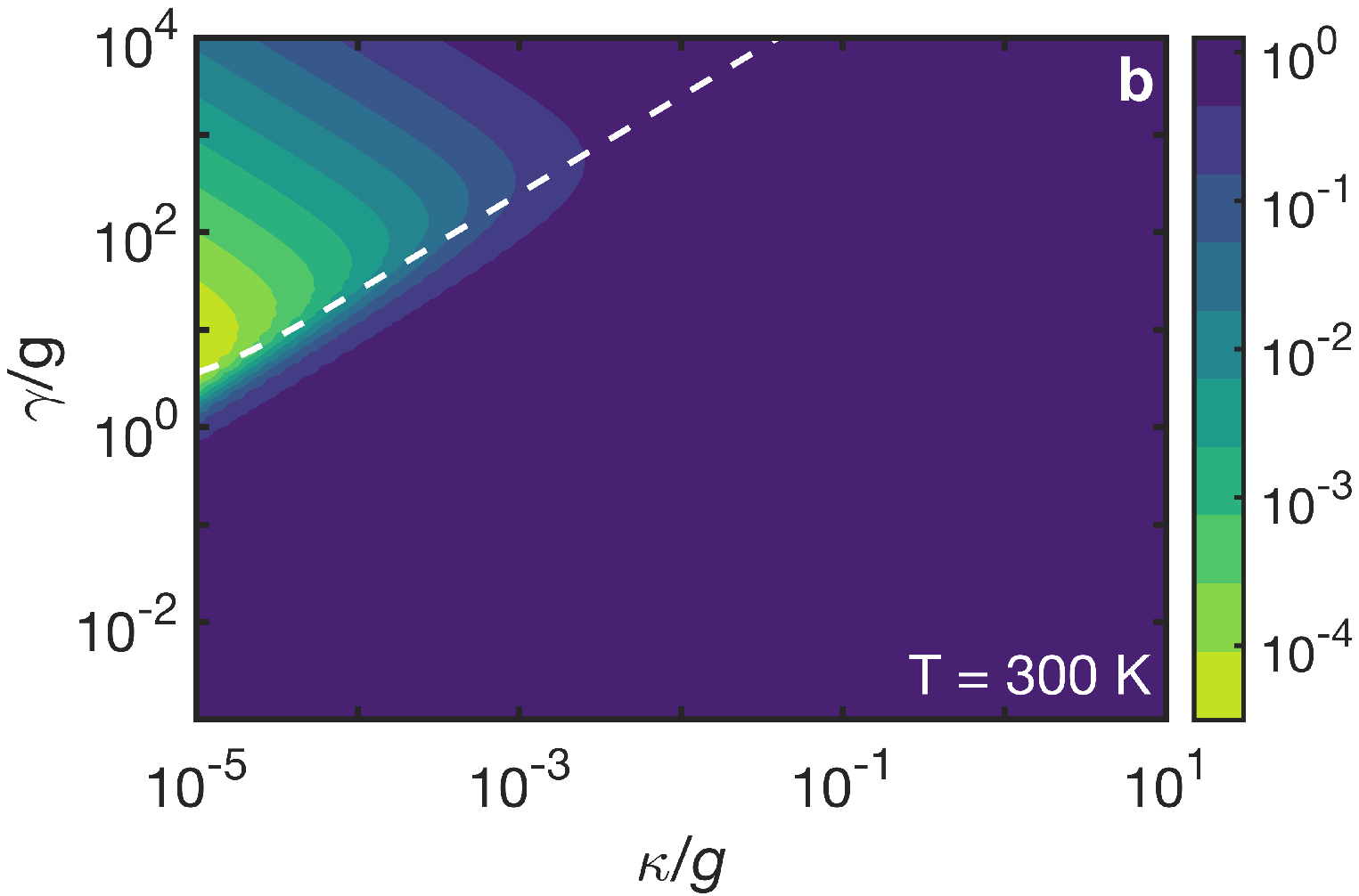}
  \end{minipage}
	\caption{Phonon cooling figure of merit, $\mathcal{F}=\braket{a^\dagger a}/n_{th}$, as a function of mechanical damping $\kappa$ and two-level system decay rate $\gamma$. Ground-state cooling is only possible in a small quadrant of the parameter space (light green color), requiring $\kappa \ll g$ and $\gamma \sim g$. The mechanical resonator ($\omega_m/2\pi = 100$ MHz) is either in : (a) a cryogenic bath, $n_{th}\sim 200$, or (b) a room temperature bath, $n_{th} \sim 6\times 10^4$.}
\end{figure*}
%%%%%%%%%%%%%%%%%%%%%%%%%%%%%%%%%%%%%%%%%%%%%%%%%%%%%%%%%%%%%%%%

\emph{Phonon number.} Following the approach from \cite{cirac1991two}, the mechanical resonator phonon number has an exact solution in the form of continued fractions,
\begin{equation}
	\braket{a^\dagger a} = n_{th} - \frac{\gamma}{2\kappa}\left(1 + \frac{a_o}{1 - \frac{a_1b_1}{1 - \frac{a_2b_2}{\cdots}}} \right)
\end{equation}
where the coefficients $a_n,b_n$ are given explicitly in appendix B. Our derivation is a generalization of \cite{cirac1991two} which includes the effects of pure dephasing. We emphasize that this solution is applicable in the weak and strong coupling regimes. For simplicity, we have ignored the effects of dephasing in the subsequent analysis. 

Using this exact solution, we now investigate the conditions required for ground-state cooling. In Figure 2, we present universal parameter plots for the phonon cooling figure of merit, $\mathcal{F} = \braket{a^\dagger a}/n_{th}$, involving the ratio of the steady-state phonon occupation number to the thermal bath number. The results are shown for a mechanical resonator, $\omega_m/(2\pi)=100$ MHz, either in a cryogenic temperature environment ($n_{th} \sim 200$) or a room temperature environment ($n_{th}\sim 6\times 10^4$), see Fig. 2-a and 2-b respectively. In both cases, ground-state cooling occurs when: (1) $\kappa\ll g$ and (2) $\gamma \gtrsim g$. The second condition corresponds to \emph{critical coupling} between the two-level system and the mechanical resonator.

\emph{Critical coupling.} To quantify the critical coupling condition required for efficient cooling, we make the mean-field approximation (see Appendix C) to obtain a tractable expression valid in our regime of interest. Assuming the coupling strength $g$ is fixed, the mechanical resonator exhibits a minimum in the phonon number, $\frac{\partial\braket{a^\dagger a}}{\partial\gamma} = 0$, when the decay rate is approximately equal to 
\begin{equation}
    \gamma/2 \approx n_{th}\kappa + \sqrt{ n_{th}^2\kappa^2 + g^2},
    \label{CriticalCouplingMean}
\end{equation}
which we refer to as the optimal cooling \emph{critical coupling} condition -- shown as dashed lines in Figure 2. We emphasize this condition as one of the primary contributions of this manuscript. Under critical coupling, the phonon number has the minimum value:
\begin{equation}
    \braket{a^\dagger a}_\text{crit} = \frac{1}{4}\frac{n_{th}\kappa }{n_{th}\kappa +\gamma/4} \left(\frac{\kappa+2\gamma}{\gamma}\right).
\end{equation}
From this expression, we find ground-state cooling occurs when $\kappa \rightarrow 0$ corresponding to mechanical resonators with very large quality factors, $Q=\omega_m/\kappa$. In this limit, the critical coupling condition takes the simple form,
\begin{equation}
 	\gamma/2 = g.
 	\label{CriticalCoupling}
\end{equation}
% cooling, specified by $\braket{a^\dagger a} \ll 1$, we make the low-excitation approximation $n_{th}\ll 1$. In this limit, we find the phonon number takes the form:
% \begin{equation}
% 	\braket{a^\dagger a} = \frac{\kappa}{\kappa + \frac{\gamma f}{\gamma + f}}n_{th} 
% \end{equation}
% where $f = \frac{4g^2}{\kappa + \gamma}$. Assuming the coupling strength $g$ is fixed, the decay rate required for maximum cooling must satisfy the condition, $\frac{\partial \braket{a^\dagger a}}{\partial\gamma} = 0$, yielding the exact critical coupling relation
% \begin{equation}
% 	g = \gamma/2.
% 	\label{CriticalCoupling}
% \end{equation}
In cavity QED, this condition is known to separate the weak and strong coupling regimes for a two-level system interacting with a single quantized mode. The weak coupling regime describes the irreversible transfer of energy from the phonon mode to the TLS, while the strong coupling regime describes an oscillatory transfer of energy between the TLS and phonon mode with Rabi frequency $g$. The critical coupling condition (\ref{CriticalCoupling}) specifies the optimal decay rate where the atom absorbs the phonon with minimal back transfer to the mechanical resonator. In passive cooling schemes, this back transfer ultimately reduces the phonon cooling figure of merit. 

We note that critical coupling is a well known concept in a wide variety of photonic applications including near-field thermophotovoltaics, microwave engineering, and optoelectronic devices based on perfect absorption \cite{yariv2000universal,piper2014total,jeon2018pareto,zanotto2014perfect,cai2000observation}. Specifically, the concept of critical coupling emerges in the design of resonators coupled to waveguides where light from the waveguide is maximally transferred to the resonator with zero back reflections. The resulting transmission at the output of the waveguide is zero. Interestingly, the concept of \emph{critical power} is also known in optomechanics to denote the maximum laser power before the onset of bistability in the effective mechanical potential \cite{aspelmeyer2014cavity}. In the phonon cooling protocol presented in this manuscript, critical coupling occurs at the Fock space level. A schematic of the corresponding energy level diagram is shown in Figure 3 (shown for 2 atoms but the overall idea is applicable to the single atom case as well). Critical coupling ensures maximum transfer from the $\ket{n}$ to $\ket{n-1}$ states with minimal back transfer to the $\ket{n}$ state, thereby ensuring optimal cooling.

% For now, we discuss how these conditions compare to realistic systems. 
In practical systems, such as NV centers in diamond, single two-level system cooperativities, $C = g^2/\kappa\gamma$, are typically much smaller than one \cite{lee2017topical}. In diamond, the spin-strain coupling strength, $g/2\pi$, is on the order of 1 Hz. State-of-the-art mechanical resonators ($\omega_m \gtrsim 1$ MHz) have quality factors on the order of $10^5-10^6$, resulting in $g\lesssim \kappa$. On the other hand, the excited-state lifetime of an NV center is about 10 ns, $\gamma/2\pi \sim 0.1$ GHz, meaning that the coupling strength is many orders of magnitude smaller than the decay rate, $g \ll \gamma$. Both of these factors explain why ground-state cooling is difficult to achieve. In the following sections, we outline two approaches for achieving critical coupling: (1)  increasing $g$ using an ensemble of non-interacting two-level systems, or (2) decreasing $\gamma$ using the subradiant eigenstates of an interacting two-level system ensemble. In the latter case, we show it is possible to increase $g$ and decrease $\gamma$ simultaneously.

% \begin{figure}[b]
% 	\centering
% 	\includegraphics[width=8cm]{FinalNvsCooperativity.jpg}	
%     \caption{Steady-state phonon occupation number as a function of cooperativity.}
% \end{figure}

%\section{Main idea}
%A general figure of merit for any cooling protocol is defined as the ratio, $\mathcal F = \frac{n_{th}}{n}$, of the initial thermal number $n_{th}$ to the final phonon number $n = \braket{a^\dagger a}$ after cooling. Assuming both the quality factor, $Q = \omega_m/\kappa$, and spin-strain coupling $g$ of the mechanical resonator is fixed, the figure of merit $\mathcal{F}$ is optimized (see Sec. 3) when the critical coupling condition
%\begin{equation}
%	g = \gamma/2
%\end{equation}
%is satisfied for a given NV center with decay rate $\gamma$. 

%While the electronic ground-state spin-strain interaction is on the order of $\sim 0.1$ MHz, recently, it was confirmed that the excited-state spin-strain interaction is an order of magnitude larger, $\sim 10$ MHz. In comparison, the linewidths of the $\ket{m_s = \pm 1}$ spin sub-levels typically lie in the $10^{-2}$ to $1$ GHz range. The implication is that the critical coupling condition may not be generally satisfied resulting in sub-optimal cooling figure of merit, $\mathcal F \approx 1$. 

\section{Cooling with non-interacting atoms}

We now consider the case of a single mechanical mode coupled to $N$ non-interacting two-level systems. This system is described by the Tavis-Cummings Hamiltonian, 
\begin{equation}
	H = \sum_i^N \frac{\
	\omega_i}{2} \sigma_{zi} + \omega_m a^\dagger a + \sum_i^N g_i(a^\dagger \sigma_i + a \sigma_i^\dagger).
\end{equation}
The open system dynamics obey the Lindblad master equation similar to Eq. (2), however, to the best of our knowledge this set of equations is not exactly solvable using the methods described above. If we assume homogeneous coupling $g_i = g$ and $\omega_i = \omega_o $, and define the collective spin operators $S_z = \tfrac{1}{2}\sum_i \sigma_{zi}$ and $S^{+} = \sum_i \sigma_i^\dagger$, we may re-write the Tavis-Cummings Hamiltonian as, $H = \omega_o S_z + \omega_m a^\dagger a + g(a^\dagger S + a S^+)$. Using the Holstein-Primakoff transformation, $S_+ = b^\dagger(N - b^\dagger b)^{1/2},\; S_- = (N - b^\dagger b)^{1/2}b,\;S_z = b^\dagger b - \frac{N}{2}$, we obtain the leading order Hamiltonian 
\begin{equation}
	H \approx \omega_o b^\dagger b + \omega_m a^\dagger a + \sqrt{N}g(a^\dagger b + a b^\dagger)
\end{equation}
describing the interaction between two harmonic oscillators with effective coupling strength $\sqrt{N}g$. This result is valid in the weak-excitation limit, $\braket{b^\dagger b}\ll N$, which is always satisfied if $n_{th} \ll N$. As shown in Appendix D, this Hamiltonian has the exact steady-state solution,
\begin{equation}
	\braket{a^\dagger a} = \frac{\kappa}{\kappa + \frac{\gamma f_N}{\gamma + f_N}}n_{th} 
\end{equation}
with $f_N = N g^2\frac{(\kappa + \gamma + 2\gamma_\phi)}{(\omega_o-\omega_m)^2 + (\kappa + \gamma + 4\gamma_\phi)^2/4}$. Assuming zero detuning, $\omega_o = \omega_m$, and zero dephasing, $\gamma_\phi = 0$, we find the critical coupling condition for an ensemble of non-interacting atoms is
\begin{equation}
    \gamma/2 = \sqrt{N}g.
\end{equation}
This analytical result confirms that increasing the number of two-level emitters allows the critical coupling condition to be fulfilled. It is worth noting that for weak coupling strengths, $g/2\pi \sim$ Hz, the number of two-level systems required for critical coupling becomes either (a) unreachable or (b) approaches a limit where interactions become unavoidable. In this regard, we will demonstrate that a suitably engineered system can utilize these interactions to its advantage by simultaneously enhancing the coupling factor $g$ while decreasing the effective decay rate $\gamma$. While the present analysis ignores the effects of individual decay processes which may lead to detrimental effects, as discussed in Ref. \cite{carmele2014opto}, we do consider the role of these decay processes in the following section for interacting two-level atoms. As we show below, the critical coupling and selective coupling conditions are important to overcome these detrimental effects.

%that there is one major drawback to this approach. By working in the weak-excitation limit $\braket{b^\dagger b}\ll N$, we have removed the inherent non-linearity of the two-level system. A quantum harmonic oscillator by its very nature is linear and cannot be saturated. This removes the ability to detect or prepare non-classical mechanical states and, depending on the application, may be detrimental to the overall goal of phonon cooling. 

%To overcome this drawback, we propose the use of interacting two-level systems to achieve critical coupling. Interestingly, while the high density limit is typically associated with detrimental interactions, we instead show that a suitably engineered system can utilize these interactions to enhance the cooling efficiency. 

%%%%%%%%%%%%%%%%%%%%%%%%%%%%%%%%%%%%%%%%%%%%%%%%%%
\begin{figure}[t!]
	\centering
    \includegraphics[width=8.5cm]{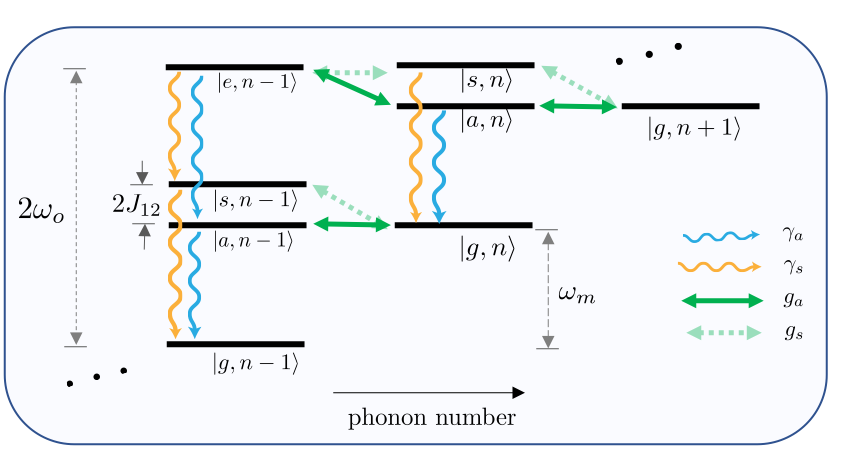}
    \caption{Energy level diagram depicting relevant transitions for resonant phonon cooling with two interacting two-level atoms. The atomic system has four eigenstates corresponding to both atoms in their ground states $\ket{g}$, both atoms in their excited states $\ket{e}$, as well as the symmetric and anti-symmetric states $\ket{s}$ and $\ket{a}$ with decay rates $\gamma_s$ and $\gamma_a$ (denoted by orange and blue arrows) which give rise to superradiance and subradiance respectively. The superradiant and subradiant states have an interaction energy splitting $2J_{12}$ and interact with the mechanical mode with coupling strengths $g_s = \tfrac{1}{\sqrt{2}}(g_1 + g_2)$ and $g_a = \tfrac{1}{\sqrt{2}}(g_1 - g_2)$. Critical coupling ($g_a \sim \gamma_a/2$) and odd-parity coupling ($g_1 = -g_2$) ensures the internal dynamics follows the ladder of states: $\ket{g,n+1}\rightarrow \ket{a,n}\rightarrow \ket{g,n}\rightarrow\ket{a,n-1}\rightarrow\ket{g,n-1}\rightarrow\cdots$, resulting in optimal cooling. In the schematic the mechanical resonator frequency $\omega_m$ is equal to the anti-symmetric state frequency, $\omega_o - J_{12}$.
}
\end{figure}
%%%%%%%%%%%%%%%%%%%%%%%%%%%%%%%%%%%%%%%%%%%%%%%%%%

%%%%%%%%%%%%%%%%%%%%%%%%%%%%%%%%%%%%%%%%%%%%%%%%%%
\begin{figure*}[t!]
	\centering
    \begin{minipage}[b]{0.495\textwidth}
    \includegraphics[width=\textwidth]{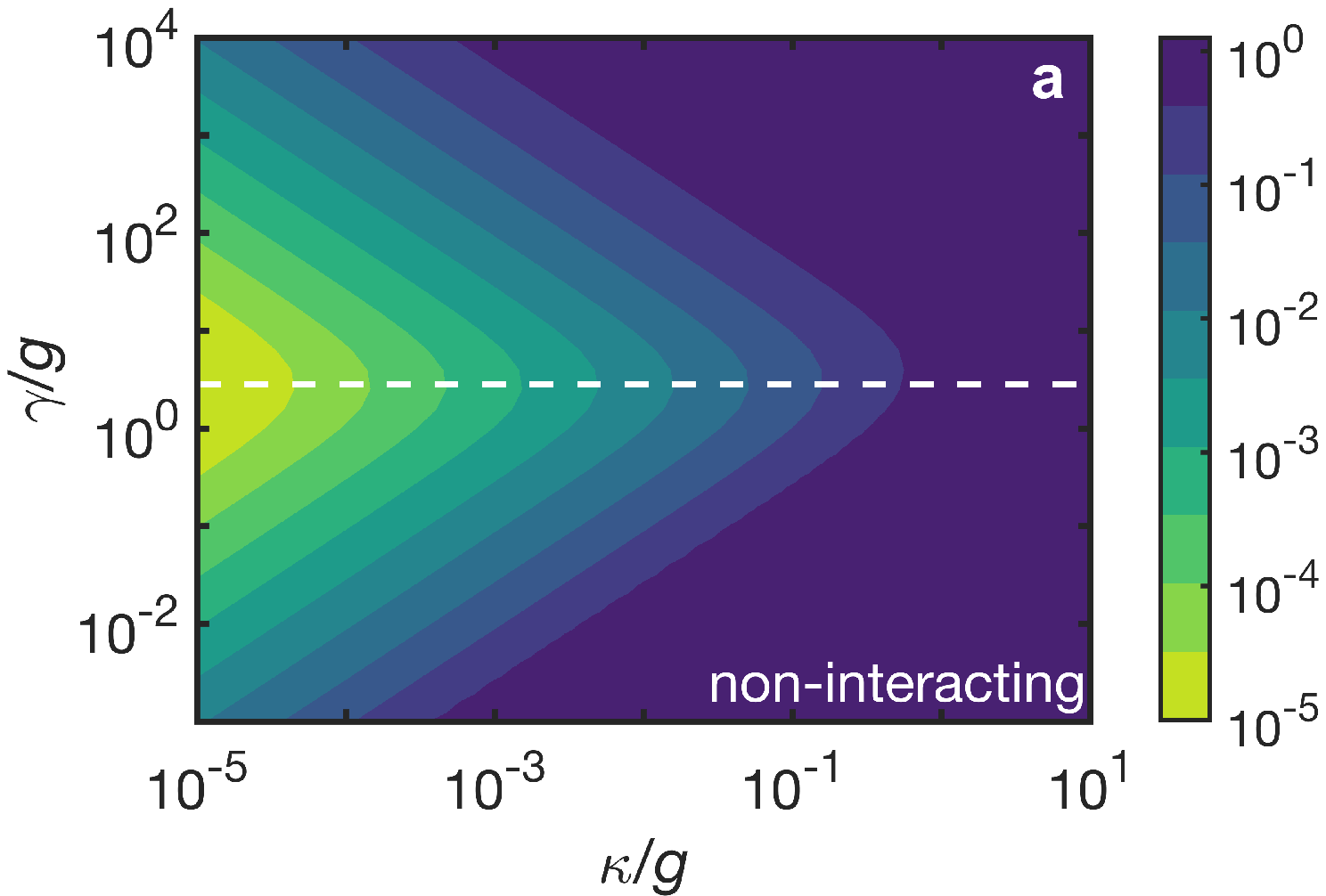}
  \end{minipage}
  \hfill
  \begin{minipage}[b]{0.49\textwidth}
    \includegraphics[width=\textwidth]{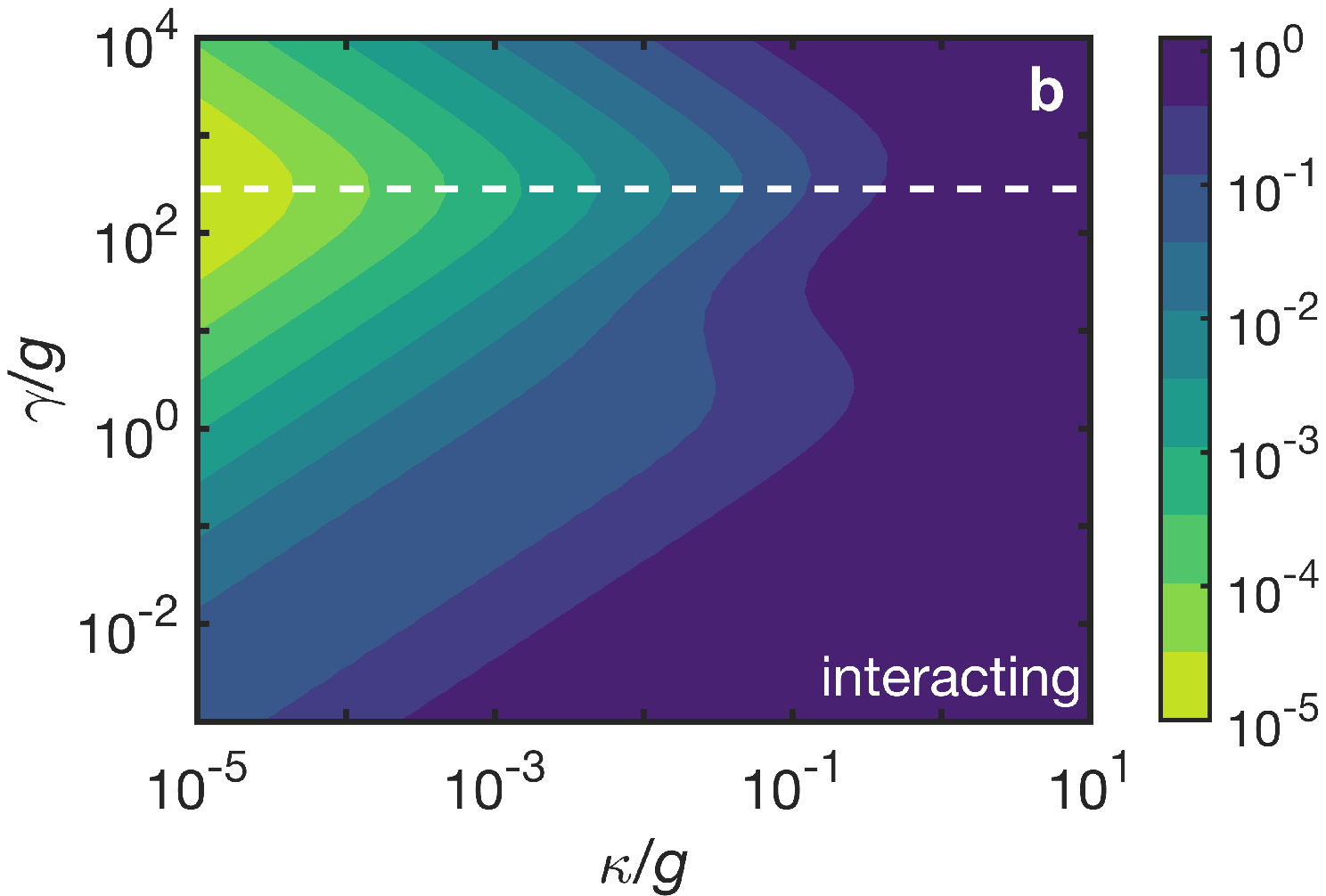}
  \end{minipage}
    \caption{Phonon cooling figure of merit $\mathcal{F}$ for (a) 2 non-interacting atoms ($J_{12} = \gamma_{12} = 0$) and (b) 2 interacting atoms ($J_{12} = 0,\; \gamma_{12} = \tfrac{99}{100}\gamma$). The optimal cooling regime (light green region) occurs when $\gamma/2 = \sqrt{2}g$ (dashed line) for non-interacting atoms, but is altered dramatically ($\gamma \gg g$) for interacting atoms ($\gamma_a/2 = \sqrt{2}g$, dashed line). We assumed anti-symmetric coupling, $g_1 = -g_2$, where the mechanical resonator is embedded in a thermal bath with $n_{th} = 2$. }
\end{figure*}
%%%%%%%%%%%%%%%%%%%%%%%%%%%%%%%%%%%%%%%%%%%%%%%%%%

\section{Cooling with interacting atoms}

In this section, we develop the quantum theory of cooling for a mechanical resonator coupled to an interacting ensemble of two-level atoms. We first consider resonant phonon cooling with two interacting two-level atoms. 

\subsection{Two interacting atoms} 
We consider 2 two-level systems that are each coupled to the resonator with coupling strengths $g_1$ and $g_2$ respectively. The total Hamiltonian is given by 
\begin{align}
    H= \!\!\sum_{i}\omega_i\sigma_i^\dagger\sigma_i+g_i(\sigma_i^\dagger a + \sigma_i a^\dagger) +J_{12}(\sigma_1^\dagger\sigma_2 + \sigma_1\sigma_2^\dagger) \nonumber 
\end{align}
where $J_{12}$ is an interaction potential between the atoms. We do not specify the origin of the interaction potential as it is highly dependent on the system in question. It may equally arise from a resonant dipole-dipole interaction between excited-state atoms, or from coupling to an optical cavity. This quantity ultimately breaks the spectral degeneracy between the single-excitation energy levels (see Figure 3), and generates many-body eigenstates with unique properties that we will explore closely in this section. We also consider the incoherent cooperative decay rate $\gamma_{ij}$ described by the non-local Lindblad superoperator,
\begin{equation}
   \sum_{i\neq j} \gamma_{ij}\mathcal{D}_{ij}(\rho) = \sum_{i\neq j}\gamma_{ij}(\sigma_j\rho\sigma_i^\dagger - \frac{1}{2}(\sigma_i^\dagger\sigma_j\rho + \rho\sigma_i^\dagger\sigma_j))
\end{equation}
which is added to the Lindblad Master equation (2). This type of Lindblad term arises when a multi-TLS system interacts with a common bath. For example, the well-known resonant dipole-dipole interaction, consisting of a coherent dipole potential $J_{ij}\propto r_{ij}^{-3}$ and cooperative decay rate $\gamma_{ij}\propto \sqrt{\gamma_i\gamma_j}$, arises when multiple atoms interact with the quantum electrodynamic vacuum \cite{ficek2002entangled,dung2002resonant,varada1992two}. Reservoir engineering techniques have also been proposed to create or modify the interaction term, $\gamma_{ij}$, using optical cavities, nanophotonic structures, as well as metamaterial systems \cite{van2013photon,mlynek2014observation,metelmann2015nonreciprocal,cortes2017super,newman2018observation,mahmoud2017dipole}.

To demonstrate how the critical coupling condition is modified due to interactions, it is sufficient to consider the dynamics of the single-excitation subspace described by the Schrodinger equation,
\begin{equation}
    i\partial_t\ket{\psi(t)} = H_{\text{eff}}\ket{\psi(t)} 
    \label{Schrodinger}
\end{equation}
where we have introduced the non-Hermitian Hamiltonian, $H_{\text{eff}} = H_{at} + H_{int}$, with the interaction Hamiltonian, $H_{int}=\sum_{i} g_i(\sigma_i^\dagger a + \sigma_i a^\dagger)$, and the atomic Hamiltonian is defined as
\begin{equation}
    H_{at} = \sum_{i,j}(\omega_{ij} - i\frac{\gamma_{ij}}{2}) \sigma_i^\dagger\sigma_j
\end{equation}
where $\omega_{ij} = J_{ij}$ for $i\neq j$. The single-excitation eigenstates of this Hamiltonian are: $\ket{\phi_s} = \frac{1}{\sqrt{2}}[\ket{e_1,g_2} + \ket{g_1,e_2}]$ and $\ket{\phi_a} = \frac{1}{\sqrt{2}}[\ket{e_1,g_2} - \ket{g_1,e_2}]$, known as the symmetric and anti-symmetric Dicke states respectively. The total single-excitation wavefunction is 
\begin{equation}
    \ket{\psi(t)} = c_g\ket{1}\otimes\ket{g_1,g_2} + \sum_k c_k\ket{0}\otimes\ket{\phi_k}.
    \label{Wavefunction}
\end{equation}
Using (\ref{Schrodinger}), we derive the equations of motion for the probability amplitudes $c_g, c_s, c_a$ for states $\ket{1,g_1,g_2}$, $\ket{0,\phi_s}$ and $\ket{0,\phi_a}$ respectively
\begin{align}
	i\dot c_g &= g_s c_s + g_a c_a  \\
    i\dot c_s &= (\omega_o+J_{12})c_s  + g_s c_g - i\frac{\gamma_s}{2}c_s 
    \label{cs} \\
    i\dot c_a &= (\omega_o-J_{12})c_a + g_a c_g - i\frac{\gamma_a}{2}c_a 
    \label{ca}
\end{align}
where $g_{s/a} = \tfrac{1}{\sqrt{2}}(g_1\pm g_2)$ and $\gamma_{s/a} = \gamma \pm \gamma_{12}$. From these equations, it is apparent that an even-parity coupling configuration ($g_1 = g_2$) will \emph{decouple} the antisymmetric state from the ground-state. On the other hand, the odd-parity coupling configuration ($g_2=-g_1$) will decouple the symmetric state from the ground state. This result shows that careful engineering of the coupling strengths $g_i$ allows cooling to proceed through specific many-body eigenstates (see Figure 3). 
%In practice, $g_i$ is a position-dependent quantity, therefore careful positional control of each atom would be required.

If we assume odd-parity coupling ($g_1 = -g_2 = g$), a straightforward calculation yields the critical coupling condition, 
\begin{equation}
    g_a = \gamma_a/2
    \label{CriticalCoupling3}
\end{equation}
valid in the high-Q limit. This condition allows for optimal cooling with realistic systems where $\gamma \gg g$.

In Figure 4, we present a full numerical simulation using QuTip for the phonon cooling figure of merit $\mathcal{F}$ for: (a) two non-interacting atoms and (b) two interacting atoms with $\gamma_{12} = \tfrac{99}{100}\gamma$. The simulations confirm the validity of Eq. (\ref{CriticalCoupling3}) showing the optimal cooling (light green region) occurs when $\gamma \gg g$ for two interacting atoms. The region of optimal cooling for non-interacting atoms satisfies, $\sqrt{2}g=\gamma/2$, and does not result in a marked cooling enhancement when $\gamma \gg g$.

Note that the interaction potential $J_{12}$ does not degrade the cooling performance. As shown in Eq. (\ref{cs}) and Eq. (\ref{ca}), the interaction potential breaks the spectral degeneracy of the symmetric and anti-symmetric states by shifting their transition frequencies to $\omega_o + J_{12}$ and $\omega_o - J_{12}$ respectively. Cooling is optimal when the frequency of the mechanical mode matches the anti-symmetric transition frequency.  

Lastly, we note that while the non-Hermitian Hamiltonian, Eq. (13) provides a qualitative description of the physics, we do find small quantitative differences when solving the full master equation (2). In the small temperature limit, $n_{th} \rightarrow 0$, these differences are small but become noticeable for higher temperatures.  For example, the small shoulder in the bottom of Figure 4-b is a direct result of coupling from the two-atom excited state $\ket{e,n}$ to the bright state $\ket{s,n}$ (see Figure 3). This shoulder becomes more noticeable at higher temperatures, but does not represent the optimal region for phonon cooling for which the non-Hermitian Hamiltonian aims to describe. 

\subsection{Critical coupling with many-body states}
For $N$ two-level systems, the general mechanism for phonon cooling is readily understood by analyzing the atomic Hamiltonian $H_\text{at}$ with right eigenstates $\ket{\phi_k}=(a_1^{(k)},a_2^{(k)},\cdots,a_N^{(k)})^T$. For convenience, we write the many-body eigenstates as a vector in the single-excitation basis $\{\ket{e_1g_2\cdots g_N},\ket{g_1e_2\cdots g_N},\cdots,\ket{g_1g_2\cdots e_N}\}$. Each eigenvector has complex eigenfrequency $\tilde{\omega}_k = \omega_k -i\gamma_k/2$. Using the Schrodinger equation, the equations of motion for the probability amplitudes are
––––––––––\begin{align}
    \dot c_g &= -i \sum_k \braket{\mathbf{g}|\phi_k} c_k \\
    \dot c_k &= -i\tilde\omega_k c_k - i \braket{\mathbf{g}|\phi_k} c_g
\end{align}
where we have introduced the coupling strength vector $\bra{\mathbf{g}} = (g_1,g_2,\cdots,g_N)$. Assuming the spontaneous emission rate for all atoms is equal to $\gamma$, the subradiant eigenstates $\ket{\phi_k}_\text{sub}$ are defined as those satisfying, $\gamma_{\text{sub}} = 2|\text{Im}\,\tilde{\omega_{k}}| < \gamma$. Selective coupling to one of the subradiant eigenstates is achieved by choosing
\begin{equation}
   \bra{\mathbf{g}} =\bra{\tilde\phi_k}_\text{sub}
   \label{SelectiveCoupling}
\end{equation}
where $\bra{\tilde\phi_k}$ is the left eigenvector of $\ket{\phi_k}$, satisfying $\braket{\tilde\phi_k|\phi_k} = \delta_{kk'}$. Note that because we are working with non-Hermitian Hamiltonians, the left eigenvector is not generally equal to the conjugate transpose of $\ket{\phi_k}$ as in Hermitian quantum mechanics. If equation (\ref{SelectiveCoupling}) is satisfied, the critical coupling condition becomes ($g_i=g$)
\begin{equation}
    \sqrt{n}g = \gamma_{\text{sub}}/2
\end{equation}
where $n\leq N$ refers to the number of atoms in the dark eigenstate $\ket{\phi_k}$. This result shows that a system with $n$-partite entanglement will simultaneously increase the effective coupling strength to $\sqrt{n}g$ while also reducing effective decay rate to $\gamma_{\text{sub}}$. In the following section, we solve the problem of a two-level system ensemble with homogeneous all-to-all interactions. This type of system arises naturally when an ensemble of atoms is confined to a spatial region that is much smaller than its transition wavelength. This limit was first studied by Dicke for a gas of molecules and is commonly referred to as the Dicke model \cite{dicke1954coherence}.

\subsection{Dicke model for cooling}
We consider a two-level system ensemble with homogeneous all-to-all interactions, $J_{ij} = J_n$ and $\gamma_{ij} = \gamma_n$, as shown in Figure 1-b. This problem has a well-known analytical solution. The eigenvalues of the non-Hermitian Hamiltonian $H_\text{at}$ consists of a single non-degenerate eigenvalue $\tilde{\omega}_1$ as well as $N-1$ degenerate eigenvalues $\tilde{\omega}_2$,
\begin{align}
     \tilde{\omega}_1 &= \omega_o + (N-1)J_n - i[\gamma + (N-1)\gamma_n]/2 \\
    \tilde{\omega}_2 &= (\omega_o - J_n)  - i(\gamma-\gamma_n)/2.
\end{align}
The first eigenvalue corresponds to the superradiant eigenstate with effective decay rate $\gamma_\text{eff}\sim N\gamma$ in the Dicke limit, $\gamma_n \rightarrow \gamma$. The eigenvector of this state is 
\begin{equation}
    \ket{\phi_1} = \tfrac{1}{\sqrt{N}}(+1,+1,\cdots,+1)^T
    \label{Wstate}
\end{equation}
which is readily recognized as the W state in quantum information science (QIS). The second eigenvalue corresponds to a many-body subradiant eigenstate with effective decay rate, $\gamma_{\text{eff}} \rightarrow 0$, in the Dicke limit $\gamma_n \rightarrow \gamma$. The eigenvectors of the degenerate subradiant subspace are not unique but must  satisfy the zero-sum relation, $a_1^{(2)} + a_2^{(2)} + \cdots a_N^{(2)} = 0$, and must be orthogonal to $\ket{\phi_1}$. These conditions allow for a large set of allowed eigenstates with either bi-partite or multi-partite entanglement. For example, it is possible to construct a set of orthogonal eigenvectors where at least one of the subradiant eigenstates has bi-partite entanglement, 
\begin{equation}
    \ket{\phi_2} = \tfrac{1}{\sqrt{2}}(+1,-1,0,\cdots,0)^T
    \label{bipartite}
\end{equation}
with the remaining eigenvectors given by a simple permutation of any two elements in (\ref{bipartite}). An orthogonal set of eigenvectors is ensured through the use of the Gram-Schmidt process, though this procedure will not ensure that all of the eigenvectors maintain bi-partite entanglement. On the other hand, it is also possible to construct a set of mutually orthogonal eigenvectors with $N$-partite entanglement. Assuming $N$ is even, a straightforward construction starts with
\begin{equation}
    \ket{\phi_2} = \tfrac{1}{\sqrt{N}}(+1,+1,\cdots,-1,-1)^T
    \label{antiW}
\end{equation}
with equal numbers of $+1$ and $-1$. We refer to this state as the \emph{anti-symmetric} W state. The remaining subradiant eigenvectors are constructed from a single permutation operation on any two elements in (\ref{antiW}). Using this construction, the entire set of eigenvectors have $N$-partite entanglement and are mutually orthogonal. We emphasize that selective coupling to either the bi-partite (\ref{bipartite}) or multi-partite entangled (\ref{antiW}) states would result in subradiance, albeit with different effective coupling strengths. 

\subsection{Strain profile engineering} 

We now discuss practical approaches for achieving the selective coupling condition (\ref{SelectiveCoupling}). This condition amounts to carefully engineering the local phase of the coupling strengths $g_i = |g_i|e^{i\phi_i}$. To achieve this level of control, we propose the use of the \emph{strain profile} in mechanical resonators. Within a material, regions that are compressed have negative strain while regions that are stretched have positive strain \cite{ovartchaiyapong2014dynamic,meesala2016enhanced,sohn2018controlling}. By embedding solid-state defects in a region displaying an odd-parity strain profile (shown in Figure 1-b), it would be possible to achieve the selective coupling condition (\ref{SelectiveCoupling}) with the anti-symmetric W state (\ref{antiW}) in the Dicke limit with homogeneous all-to-all interaction. When this condition is satisfied, the cooling protocol will proceed through the dark (subradiant) entangled states of the interacting two-level system ensemble resulting in efficient cooling. 

% If we consider the $N=4$ case explicitly, the subradiant subspace may be described by the multi-partite eigenvectors,
% \begin{align}
%     \ket{\phi_2} =  \tfrac{1}{2}(+1,-1,+1,-1) \\
%     \ket{\phi_3} =  \tfrac{1}{2}(+1,+1,-1,-1) \\
%     \ket{\phi_4} =  \tfrac{1}{2}(+1,-1,-1,+1) 
% \end{align}
% or bipartite eigenvectors,
% \begin{align}
%     \ket{\phi_2} &=  \tfrac{1}{\sqrt{2}}(+1,-1,0,0) \\
%     \ket{\phi_3} &=  \tfrac{1}{\sqrt{2}}(+1,0,-1,0) \\
%     \ket{\phi_4} &=  \tfrac{1}{\sqrt{2}}(+1,0,0,-1). 
% \end{align}

\section{Discussion} 

We now discuss the role of {entanglement}. At first glance, it seems like entanglement should not be necessary to explain the collective phenomena considered in this manuscript. This is corroborated by the observation that collective effects, such as the $\sqrt{N}g$ collective coupling dependence, are also known to appear in classical physics. For example, consider $N$ classical oscillators coupled to a single damped harmonic oscillator. It is possible to show that the effective decay rate of the damped oscillator scales as $\sim Ng^2$. The classical collective effect occurs because the initial energy $\mathcal{E}_i$ is divided into $N$ partitions, each with energy $\mathcal{E}_i/N$. In the quantum regime, this cannot occur. A single quantum of energy cannot be divided into smaller components when considering systems with linear interactions. As a result, the single quantum must be shared among $N$ systems, see Eq. (25) and Eq. (27). This implies entanglement is required to explain the cooperative effects highlighted in this paper. We must emphasize that these arguments are specifically relevant for systems with {distinguishable} particles. As pointed out in \cite{wolfe2014certifying}, the permutational symmetry of a collection of {indistinguishable} particles removes entanglement by inducing separability in the many-body wavefunction. Entanglement is not required to explain superradiance for an ensemble of indistinguishable two-level systems. 

%From a QIS perspective, it is worth asking whether entanglement explains the collective properties of the two-level system ensemble. 

\subsection{Entanglement properties} 
From a QIS perspective, it is also worth asking whether \emph{specific} multi-partite entangled states are better suited for phonon cooling. In this regard, comparing the GHZ and W states, representing two inequivalent forms of multi-partite entanglement, allows us to provide a qualitative answer to this question. %While the $W$ state retains maximally bipartite entanglement when any of the two-level systems are traced out, the $GHZ$ state

The GHZ state, $\ket{GHZ} = \tfrac{1}{\sqrt{2}}(\ket{g}^{\otimes N} + \ket{e}^{\otimes N})$, describes a superposition of all atoms in their ground state with all atoms in their excited-state. As $N$ atoms participate in this state, one expects an effective coupling strength scaling of $\sqrt{N}g$. Similarly, the all-excited state $\ket{1}^{\otimes N}$ will have an effective decay rate that will scale with the number of atoms $\sim N\gamma$. These two properties imply the critical coupling condition is unreachable assuming starting parameters $\gamma \gg g$. This makes the GHZ state an unsuitable candidate for phonon cooling.

On the other hand, the W state, Eq. (\ref{Wstate}), describes a multi-partite system in the single-excitation superposition state. Similar to the GHZ state, the effective coupling strength should scale as $\sqrt{N}g$. In contrast to the GHZ state, the single excitation nature of the W state for non-interacting atoms suggests an effective decay rate scaling of $\sim \gamma$. The W state is therefore a suitable candidate for achieving critical coupling. 

Indeed, our work provides quantitative confirmation of this qualititative argument. The Tavis-Cummings Hamiltonian is a representative model of $N$ non-interacting atoms. The $W$ state is one the single-excitation eigenstates of this Hamiltonian \cite{otten2016origins,otten2015entanglement}, assuming homogeneous coupling $g_i = g$. Using Eqs. (\ref{Schrodinger})-(\ref{Wavefunction}), it is possible to confirm the W state has effective coupling strength $\sqrt{N}g$ and decay rate $\gamma$ in agreement with the arguments given above. Note that in addition to the W state, all other single-excitation superposition states would also share the same properties according to the Tavis-Cummings model in the large number limit. 

Taking into account the role of interactions, namely the cooperative decay rate $\gamma_{ij}$, we found two distinct W-like states. First, the symmetric W state, Eq. (\ref{Wstate}), with effective decay rate $\gamma_\text{eff}\rightarrow N\gamma$. Second, the anti-symmetric W state, Eq. (\ref{antiW}), with effective decay rate $\gamma_\text{eff}\rightarrow 0$ in the Dicke limit. Since both states have an effective coupling strength scaling of $\sqrt{N}g$, this suggests the anti-symmetric $W$ state is among the most ideal multi-partite entangled state for optimal phonon cooling. 

\subsection{Practical implementations} 
There exists a wide variety of quantum hybrid systems combining nanomechanical resonators with two-level systems. The two-level system may represent superconducting (SC) qubits, ultracold atoms or ions, as well as solid-state defects or quantum dots. Table I provides an order of magnitude summary for the relevant empirical parameters for these systems \cite{lee2017topical,macquarrie2013mechanical,macquarrie2017cooling}. There is a clear disparity ($g \ll \gamma$) between the single two-level system coupling factor $g$ and the decay rate $\gamma$ for all  experimental systems. This imbalance highlights the necessity for engineered systems with increased coupling strengths and suppressed decay rates. %Note that both superconducting qubits and ultra-cold atoms have a much smaller discrepancy between $g$ and $\gamma$. 
Note that the disparity between the spin-strain coupling $g$ and decay rate $\gamma$ for NV centers is as large as six orders of magnitude. Our proposal allows for the realization of ground-state cooling overcoming this obstacle. 

\begin{table}[h!]
\caption{Empirical parameters from \cite{lee2017topical}.}
\begin{center}
\begin{tabular}{|l|c|c|}
    \;\;\;\;\;\;\;\;\;\;\;\;\textbf{Platform} & $g/2\pi$   & $\gamma/2\pi$  \\ \hline
    SC device (capacitive) & $\sim 1$-100 MHz & $\sim 0.01$ - 1 GHz \\   \hline
    SC device (inductive) & $\sim 10$ kHz & $\sim 1$ MHz \\   \hline
    Cold atoms/ions & $\sim $ kHz & $\sim 10$ kHz \\   \hline
    NV centers (spin) & $\sim$ Hz & $\sim$ kHz to  MHz  \\ \hline
    NV centers (orbital) & $\sim$ kHz & $\sim$ GHz \\ \hline
\end{tabular}
\vspace{0.1cm}
\vspace{0.1cm}
\\ $^\dagger$ $\gamma$ denotes dominant dephasing or decay rate.
\end{center}
\end{table}

\subsection{Effect of disorder} 
Finally, it is worth considering whether static disorder and dephasing have detrimental effects for phonon cooling based on many-body subradiance. As highlighted in \cite{temnov2005superradiance}, an ensemble of $N$ inhomogeneously broadened two-level systems coupled to a low-Q cavity exhibits bi-exponential emission dynamics corresponding to superradiant and subradiant decay. This illustrates that subradiance may persist in the presence of static disorder and dephasing. While the system in \cite{temnov2005superradiance} seems qualitatively different than our system based on interacting two-level systems (interaction given by $J_{ij}$ and $\gamma_{ij}$), we re-emphasize that these terms arise when $N$ two-level systems interact with a common bath. By tracing out the optical cavity in the referenced study, one may derive an effective Hamiltonian that agrees with the one considered in this manuscript. To this end, we expect the general idea of our phonon cooling proposal to succeed in the presence of static and dynamic disorder.

%The critical coupling conditions that have been discussed throughout this manuscript are derived in the high-Q limit where $\kappa \rightarrow 0$. However, as is clearly seen in Figure 2\footnote{As well as figure 3, but it is less clear because we have assumed a low temperature configuration where $n_{th} = 2$}, the critical coupling condition for finite values of $\kappa$ in the high temperature limit take the approximate form $\gamma/2 = \sqrt{n_{th}+1}g$, which may be proven using the exact solution from section 1. 

 %Finally, from a quantum information science perspective, we should mention the connection between multipartite entanglement and subradiance 

% \begin{figure}
% 	\centering
%     \includegraphics[width=15cm]{2AtomJCladder.png}
% \end{figure}
\section{Conclusion}

To summarize, we have presented a general open quantum systems model for ground-state cooling based on critical coupling. By engineering the atom-resonator coupling parameters using the strain profile of the mechanical resonator, we illustrated how resonant phonon cooling proceeds through the dark entangled states of the two-level system ensemble. Ultimately, this process enables ground-state cooling under the weak coupling condition, $\gamma \gg g$. We also discussed the role of entanglement and highlighted cooperative effects as a key factor in enhancing the cooling figure of merit. We emphasize that our results are universal to a wide variety of systems including silicon and nitrogen vacancy centers in diamond as well as quantum dots. Our results pave the way for ground-state cooling experiments using solid-state defects in the near future.

\vspace{2cm}

\section*{Appendix A: 3-level system description}

As noted in the introduction, the Hamiltonian describing phonon cooling with a three-level system is
\begin{equation}
    H = \omega_p \sigma_{gg} + (\omega_p + \omega_o)\sigma_{ee} + \omega_m a^\dagger a + g(a^\dagger  \sigma_{ge} + a \sigma_{ge}^\dagger).
\end{equation}
We have set the frequency of the $\ket{p}$ state to zero without loss of generality. We have also defined the generalized Pauli operators $\sigma_{nm} = \ket{n}\!\bra{m}$. The three-level open quantum system obeys the master equation,
\begin{align}
	\dot\rho &= -i[H,\rho] + \gamma_g \mathcal{D}[\sigma_{pg}]\rho + \gamma_e \mathcal{D}[\sigma_{pe}]\rho+ P\mathcal{D}(\sigma_{pg}^\dagger)\rho  \nonumber \\
	&+ \kappa(n_{th}+1)\mathcal{D}[a]\rho + \kappa n_{th}\mathcal{D}[a^\dagger]\rho
\end{align}
where we have included the effect of pumping through the incoherent Lindblad term $\mathcal{D}(\sigma_{pg}^\dagger)\rho$ with pump rate $P$. The decay rates $\gamma_g$ and $\gamma_e$ refer to the transitions $\ket{g}\rightarrow \ket{p}$ and $\ket{e}\rightarrow \ket{p}$ respectively. Using these expressions, we derive the Heisenberg equations of motion for states $\ket{p},\ket{g}$ and $\ket{e}$,
\begin{align}
    \partial_t\braket{\sigma_{pp}} &= -P\braket{\sigma_{pp}} +\gamma_g\braket{\sigma_{gg}} + \gamma_e\braket{\sigma_{ee}}  \\
    \partial_t\braket{\sigma_{gg}} &=  - ig\braket{a^\dagger \sigma_{ge} - a\sigma_{ge}^\dagger } -\gamma_g\braket{\sigma_{gg}} +P\braket{\sigma_{pp}} \nonumber  \\
    \partial_t\braket{\sigma_{ee}} &=  + ig\braket{a^\dagger \sigma_{ge} - a\sigma_{ge}^\dagger } - \gamma_e\braket{\sigma_{ee}} \nonumber\\
    \partial_t\braket{\sigma_{ge}} &=  + ig\braket{a(\sigma_{ee} - \sigma_{gg}) }  -(i\omega_o+\frac{\gamma_g + \gamma_e}{2})\braket{\sigma_{ge}} \nonumber
\end{align}
To obtain a closed set of equations for the two-level subspace $\ket{g}$ and $\ket{e}$, we use the relation,
$\braket{\sigma_{pp}}+\braket{\sigma_{gg}}+\braket{\sigma_{ee}} = 1$, together with the fast pump approximation, $\partial_t \braket{\sigma_{pp}} \approx 0$, valid in the limit $P \gg \gamma_e,\gamma_g$, giving
\begin{align}
    \partial_t\braket{\sigma_{gg}} &\approx  - ig\braket{a^\dagger \sigma_{ge} - a\sigma_{ge}^\dagger }  +\gamma_e\braket{\sigma_{ee}}   \\
    \partial_t\braket{\sigma_{ee}} &\approx  + ig\braket{a^\dagger \sigma_{ge} - a\sigma_{ge}^\dagger } - \gamma_e\braket{\sigma_{ee}} \nonumber\\
    \partial_t\braket{\sigma_{ge}} &\approx   ig\braket{a(2\sigma_{ee} - 1) } - (i\omega_o+\frac{\gamma_g + \gamma_e}{2})\braket{\sigma_{ge}} \nonumber
\end{align}
These equations correspond closely with the equivalent Heisenberg equations of motion for the two-level system Hamiltonian (1) using the Lindblad Master equation (2). In the steady-state limit, the three-level system description is exactly equivalent to the two-level system results when $P \gg \gamma_e \gg \gamma_g$. For the case of equal decay rates, $\gamma = \gamma_e = \gamma_g$, the two-level system description requires an additional dephasing rate, $\gamma_\phi = \gamma$, captured by the Lindblad term, $\gamma_\phi\mathcal{D}[\sigma_{ee}]\rho$, in order to exactly match the three-level system description when $P \gg \gamma$.

In the limit of coherent pumping, the conditions for recovering an effective two-level system description are more subtle. Our simulations confirm that a coherent pump with zero detuning and intermediate pump power correspond closely with the two-level system results, albeit with an overall smaller phonon cooling figure of merit. In general, the detuning and pump power will affect the critical coupling condition. Further details will be included in future work. 

%We should note, however, that in the limit of coherent pumping, explicitly given by an additional Hamiltonian term, $H_p = \Omega_p(\sigma_{pg} e^{i\omega_pt} + \sigma_{pg}^\dagger e^{-i\omega_pt})$, this exact equivalence does not always hold. Under strong atom-field coupling conditions, the dressed states of the atom and field result in a shift in the frequency for the $\ket{g}$ state, resulting in a modification for $\omega_o$. 

\section*{Appendix B: exact solution to Eq. (2)}

Following the approach from \cite{cirac1991two}, we present an exact solution to equations (1) and (2) in the manuscript. The presented solution includes the effect of dephasing and is a generalization of the results in \cite{cirac1991two}. We first define the normal-ordered characteristic function,
\begin{equation}
	C(\sigma^j,t) = e^{\lambda_1\lambda_2 n_{th}}\text{tr}[e^{i\lambda_1 a^\dagger} e^{i\lambda_2 a} \sigma^j \rho(t) ]
\end{equation}
where $\sigma^j =\left\{1,\sigma_z,\sigma^+,\sigma  \right\}$ and $j =\left\{0,1,2,3 \right\}$.  Knowledge of the characteristic function gives all of the information about the system. Using the Glauber-Sudarshan representation $P^j(\alpha)$ for the operator $\sigma^j\rho$, the characteristic function may be written as 
\begin{equation}
    C(\sigma^j,t) = \sum_{nm} c_{n,m}^j(t)\lambda_1^{n} \lambda_2^{m}  \label{Cexpansion}
\end{equation}
where the coefficients are
\begin{equation}
    c_{n,m}^j(t) = \frac{n_{th}^n }{m!} \int\! d^2\alpha\;  P^j(\alpha,t) (i\alpha)^{m-n}   L_n^{m-n}\left(\frac{|\alpha|^2}{n_{th}}\right) 
\end{equation}
when $m\geq n$. $L_n^{k}(x)$ denotes the associated Laguerre polynomials. Using the optical equivalence theorem, the coefficients are also equal to 
\begin{equation}
    c_{n,m}^j(t) = \frac{n_{th}^n }{m!} \braket{:(ia)^{m-n}   L_n^{m-n}\left(\frac{a^\dagger a}{n_{th}}\right):\sigma^j}
\end{equation}
where $\braket{:\,:}$ corresponds to the normal-order mean value. Using the Lindblad master equation, we derive the evolution equation for the characteristic function,
\begin{widetext}
\begin{align}
\partial_t C(\sigma^j)  &= i \frac{\Delta}{2} C \left( \left[ \sigma^j , \sigma^z\right] \right) - g \left[ \left( \frac {\partial}{\partial\lambda_2} - n_{th}\lambda_1 \right) C\left(\left[ \sigma^j, \sigma^+ \right] \right) + \left(\frac{\partial}{\partial\lambda_1} - n_{th} \lambda_2 \right) C \left( \left[\sigma^j, \sigma^- \right] \right)\right]    \nonumber \\ 
&-  g \Big[ \lambda_1 C \left( \sigma^+ \sigma^j \right) - \lambda_2 C \left( \sigma^j \sigma^- \right) \Big] - \kappa \left( \lambda_{1} \frac{\partial }{\partial \lambda_1} + \lambda_2\frac {\partial }{\partial \lambda_{2} } \right) C \left(\sigma^{j} \right) \nonumber \\
&+ \frac{\gamma}{2} C \left( \left[ \sigma^{+} , \sigma^{j} \right] \sigma^{-} + \sigma^{+} \left[\sigma^{j} , \sigma^{-} \right] \right) +  \gamma_\phi C \left( \sigma_z \sigma^{j} \sigma_z - \sigma^j \right) \label{PDE1}.
\end{align}
\end{widetext}
Substituting expansion (\ref{Cexpansion}) into the partial differential equation (\ref{PDE1}) and comparing coefficients in powers of $\lambda_{1,2}$, we obtain
\begin{align}
    \dot c_{n,m}^0 &= -g( c_{n-1,m}^2 - c_{n,m-1}^3 ) - \frac{\kappa}{2}(n+m) c_{n,m}^0 \nonumber \\
     \dot c_{n,m}^1 &= 2g(n_{th}+\tfrac{1}{2})(c_{n-1,m}^2 - c_{n,m-1}^3) - 2g[(m+1)c_{n,m+1}^2 \nonumber \\ &-(n+1)c_{n+1,m}^3] - \gamma(c_{n,m}^0+c_{n,m}^1) -\frac{\kappa}{2}(n+m) c_{n,m}^1 \nonumber \\
    \dot c_{n,m}^2 &= g(n_{th} +\tfrac{1}{2})c_{n,m-1}^1 - g(n+1)c_{n+1,m}^1 + \tfrac{1}{2}g c_{n,m-1}^0 \nonumber \\ 
    &- D_{n,m} c_{n,m}^2 \nonumber \\
    \dot c_{n,m}^3 &= -g(n_{th}+\tfrac{1}{2})c_{n-1,m}^1 + g(m+1)c_{n,m+1}^1 \nonumber \\ 
    &- \tfrac{1}{2}g c_{n-1,m}^0 - D_{n,m}^* c_{n,m}^3 \nonumber 
\end{align}
where $D_{n,m} = 2\gamma_\phi + \frac{ \gamma}{2} + \frac{\kappa}{2}(n+m) + i\Delta$. In the steady-state limit, we obtain a linear system of equations for coefficients $c_{n,m}^j$ with nearest-neighbor coupling through the $n$ and $m$ indices. This property allows for the exact solution to be found in terms of continued fractions. Note that deriving the Heisenberg equations of motion for the normal-ordered function $\braket{a^{\dagger n}a^m \sigma^j}$ does not give such simple form. In the steady-state, we obtain the following set of equations,
\begin{align}
    -g P_{n-1} &= n\kappa I_n \nonumber \\
    g(2n_{th} + 1) P_{n-1} &= 2g(n+1)P_n + (\gamma + n \kappa ) W_n + \gamma I_n  \nonumber \\
    -i\Delta P_n &=  \tfrac{1}{2}\left(4\gamma_\phi +\gamma + \kappa(2n+1 ) \right) Q_n \nonumber \\
    g(2n_{th} + 1) W_n &= 2g(n + 1)W_{n+1} - g I_n +i\Delta Q_n \nonumber \\
    &+ \tfrac{1}{2}\left(4\gamma_\phi + \gamma + \kappa(2n+1 ) \right) P_n    
\end{align}
where $I_0 = 1$. Eliminating $I_n$ and $Q_n$ we obtain
\begin{align}
    T_n P_{n-1} - U_n P_n - (\gamma + n\kappa)W_n  &= 0  \nonumber \\
    L_n \mathcal{J} P_{n-1} - g(2n_{th}+1)\mathcal{J}_n W_n + U_n\mathcal{J}_n W_{n+1} + P_n &= 0 \nonumber
\end{align}
where $\mathcal{J}_n = \frac{( \gamma + \kappa(2n+1) + 4\gamma_\phi)/2}{\Delta^2 + ( \gamma + \kappa(2n+1) + 4\gamma_\phi)^2/4 }$. Similarly, we have defined ($n\geq 1$)
\begin{align*}
    U_n &= 2g(n+1)  \\ 
    L_n &= \frac{g^2}{n\kappa} \\ 
    T_n &= g\left[ 2 n_{th} + 1 + \frac { \gamma } { n \kappa }  \right] 
\end{align*} 
After extensive algebra, the final expression for the phonon occupation number, Eqn. (3), is written in terms of coefficients:
\begin{align}
 a_n &= \frac{T_n x_{n-1}} { T_n y_{n-1} + U_{n} x_n + \gamma +  n\kappa }  \\ 
 b_{n+1} &= \frac{U_{n} y_{n}}{T_n y_{n-1} + U_n x_n + \gamma +  n \kappa }  
\end{align}
where
\begin{align} 
 y_n &= \frac{T_n U_n }{ R_n }  \\  
 x_n &= \frac{g(2 n_{th} + 1)T_n - L_n( \gamma +  n \kappa ) }{R_n}  \\  
 R_n &= U_n L_n + T_n /\mathcal{J}_n
\end{align}
For $n=0$, we obtain the coefficients,
\begin{align}
    a_o &= -\frac{f+\gamma}{\gamma + f(2n_{th} + 1)} \\
    b_1 &= \frac{2f}{\gamma + f(2n_{th} + 1)}
\end{align}
where $f = 2g^2 \mathcal{J}_0$.
%The exact solution to this problem is readily obtained by introducing the X and P operators, . 

\section*{Appendix C: mean-field approximation} 
In this section, we utilize the mean-field approximation to obtain an analytically tractable expression for the phonon number $\braket{a^\dagger a}$. We compare the results to the exact solution given above written in terms of continued fractions. Using the factorization approximation, $\braket{a^\dagger a \sigma_z} \approx \braket{a^\dagger a}\braket{\sigma_z}$, we find the following \emph{closed} set of Heisenberg equations of motion,
\begin{align}
 	\partial_t \braket{a^\dagger a} &= -ig(\braket{a^\dagger \sigma} - \braket{a \sigma^\dagger}) - \kappa(\braket{a^\dagger a} - n_{th}) \\
    \partial_t \braket{\sigma_z } &= 2ig(\braket{a^\dagger \sigma} - \braket{a \sigma^\dagger}) - \gamma(1+\braket{\sigma_z}) \nonumber \\
    \partial_t \braket{a^\dagger \sigma} &\approx ig(\braket{a^\dagger a}\braket{\sigma_z} + \tfrac{1}{2}(1+\braket{\sigma_z})) - \frac{1}{2}(\kappa + \gamma) \braket{a^\dagger \sigma}. \nonumber
\end{align}
In the steady-state limit, solving this set of nonlinear equations yields the final phonon number:
\begin{align}
    \braket{a^\dagger a} &= \frac{2 n_{th} - \gamma/\kappa - 1}{4} \nonumber \\  
    &+ \left[ \frac{(1+\gamma/\kappa - 2 n_{th})^2}{16} + \frac{n_{th}(\gamma+f)}{2f} \right]^{1/2} 
\end{align}
The critical coupling condition is then given by
\begin{equation}
    g^2 \approx \frac{\gamma^2}{4} + n_{th}\gamma\kappa\left(1+\frac{2\kappa}{4n_{th}\kappa + \gamma}\right)
\end{equation}
Note the first term is independent of the thermal occupation number. Alternatively, the result may be written as
\begin{equation}
    \gamma/2 = n_{th}\kappa + \sqrt{ n_{th}^2\kappa^2 + g^2}.
\end{equation}
When the system is critically coupled, the phonon number becomes:
\begin{equation}
    \braket{a^\dagger a}_c = \frac{n_{th}\kappa (\kappa + 2\gamma)}{g^2} = \frac{1}{4} \frac{n_{th}\kappa }{n_{th}\kappa + \gamma/4}\frac{\kappa+2\gamma}{\gamma}.
\end{equation}

\section*{Appendix D: Theory for 2 quantum oscillators}
In the following, we derive the steady-state solutions for two interaction oscillators where one is coupled to a thermal bath, as described by the Lindblad Master equation (2). We consider the Hamiltonian for two quantum oscillators,
\begin{equation}
	H = \omega_o a^\dagger a + \omega_1 b^\dagger b + g(a^\dagger b + a b^\dagger)
 \end{equation}
\newline
where the operators satisfy the commutation relations, $[a,a^\dagger] = 1$, $[b,b^\dagger] = 1$, $[a,b^\dagger] = 0$. The equations of motion for the second order moments are given by
\begin{align}
	\partial_t \braket{a^\dagger a} &= -ig(\braket{a^\dagger b} - \braket{a b^\dagger}) - \kappa(\braket{a^\dagger a} - n_{th}) \\
   	\partial_t \braket{b^\dagger b} &= +ig(\braket{a^\dagger b} - \braket{a b^\dagger}) - \gamma\braket{b^\dagger b} \\
    \partial_t \braket{a^\dagger b} &= ig(\braket{b^\dagger b} -\braket{a^\dagger a}) - D_{o} \braket{a^\dagger b}.
\end{align}
where $D_o = -i(\omega_o - \omega_1) +\frac{1}{2}(\kappa + \gamma) + 2\gamma_\phi$. We emphasize that these equations include the effects of detuning and dephasing. At steady-state, the final expression for the phonon occupation number is
\begin{equation}
	\braket{a^\dagger a} = \frac{\kappa}{\kappa + \frac{\gamma f}{\gamma + f}}n_{th}
\end{equation}
where $f = g^2\frac{(\kappa + \gamma + 4\gamma_\phi)}{(\omega_o - \omega_1)^2 + (\kappa + \gamma + 4\gamma_\phi)^2/4}$. These results are used for the analysis of phonon cooling with a non-interacting ensemble. These equations are also valid for interacting systems under the low-excitation approximation. 

\subsection*{Low excitation approximation}
Ground-state cooling occurs when $\braket{a^\dagger a} \ll 1$. The conditions required for optimal cooling can therefore be understood in the weak-excitation limit, $n_{th}\ll 1$. Using the mean field approximation $\braket{f(a^\dagger,a)\sigma_z} = \braket{f(a^\dagger a)}\braket{\sigma_z}$, along with the low-excitation approximation $\braket{\sigma_z} \approx -1$, the Heisenberg equations of motion form a \emph{closed} set of coupled differential equations which are formally equivalent to the equations of motion for two coupled quantum oscillators as derived above. We use this approximation and the corresponding equations for the analysis of the critical coupling condition throughout the manuscript.

\section*{Appendix E: numerical simulations} 
For all numerical simulations involving $N>1$ two-level systems, we used the python package QuTip. The Fock state Hilbert space was truncated at a number much larger than the thermal number $n_\text{max} \gg n_{th}$. Numerical stability was confirmed by verifying the results did not change when $n_\text{max}$ was increased.

\section*{Acknowledgements}
We thank Greg Fuchs for insightful comments during the preparation of this manuscript. This work was performed at the Center for Nanoscale Materials, a U.S. Department of Energy Office of Science User Facility, and supported by the U.S. Department of Energy, Office of Science, under Contract No. DE-AC02-06CH11357.

% \paragraph{Resonant dipole-dipole interaction.} The interaction terms  $J_{ij}$ and $\gamma_{ij}$ have exact expressions if their origin is the resonant dipole-dipole interaction. They take the form, $J_{ij} = \tfrac{3}{4}\gamma g(\omega_o r/c)$ and $\gamma_{ij} = \tfrac{3}{2}\gamma f(\omega_o r/c)$ where
% \begin{align}
%     f(x) &= \alpha\frac{\sin x}{x} + \beta\left( \frac{\cos x }{x^2} - \frac{\sin x }{x^3} \right) \\
%     g(x) &= - \alpha\frac{\cos x }{x} + \beta \left(\frac{\sin x }{x^2} + \frac{\cos x }{x^3} \right)
% \end{align}
% where $\alpha = 1 - \cos^2\theta$ and $\beta = 1 - 3\cos^2\theta$, and $\theta$ represents the angle between the position vector $\mathbf{r}$ and a common dipole orientation vector.

%\bibliography{PhononCooling.bib}

\begin{thebibliography}{44}%
\makeatletter
\providecommand \@ifxundefined [1]{%
 \@ifx{#1\undefined}
}%
\providecommand \@ifnum [1]{%
 \ifnum #1\expandafter \@firstoftwo
 \else \expandafter \@secondoftwo
 \fi
}%
\providecommand \@ifx [1]{%
 \ifx #1\expandafter \@firstoftwo
 \else \expandafter \@secondoftwo
 \fi
}%
\providecommand \natexlab [1]{#1}%
\providecommand \enquote  [1]{``#1''}%
\providecommand \bibnamefont  [1]{#1}%
\providecommand \bibfnamefont [1]{#1}%
\providecommand \citenamefont [1]{#1}%
\providecommand \href@noop [0]{\@secondoftwo}%
\providecommand \href [0]{\begingroup \@sanitize@url \@href}%
\providecommand \@href[1]{\@@startlink{#1}\@@href}%
\providecommand \@@href[1]{\endgroup#1\@@endlink}%
\providecommand \@sanitize@url [0]{\catcode `\\12\catcode `\$12\catcode
  `\&12\catcode `\#12\catcode `\^12\catcode `\_12\catcode `\%12\relax}%
\providecommand \@@startlink[1]{}%
\providecommand \@@endlink[0]{}%
\providecommand \url  [0]{\begingroup\@sanitize@url \@url }%
\providecommand \@url [1]{\endgroup\@href {#1}{\urlprefix }}%
\providecommand \urlprefix  [0]{URL }%
\providecommand \Eprint [0]{\href }%
\providecommand \doibase [0]{http://dx.doi.org/}%
\providecommand \selectlanguage [0]{\@gobble}%
\providecommand \bibinfo  [0]{\@secondoftwo}%
\providecommand \bibfield  [0]{\@secondoftwo}%
\providecommand \translation [1]{[#1]}%
\providecommand \BibitemOpen [0]{}%
\providecommand \bibitemStop [0]{}%
\providecommand \bibitemNoStop [0]{.\EOS\space}%
\providecommand \EOS [0]{\spacefactor3000\relax}%
\providecommand \BibitemShut  [1]{\csname bibitem#1\endcsname}%
\let\auto@bib@innerbib\@empty
%</preamble>
\bibitem [{\citenamefont {Jost}\ \emph {et~al.}(2009)\citenamefont {Jost},
  \citenamefont {Home}, \citenamefont {Amini}, \citenamefont {Hanneke},
  \citenamefont {Ozeri}, \citenamefont {Langer}, \citenamefont {Bollinger},
  \citenamefont {Leibfried},\ and\ \citenamefont
  {Wineland}}]{jost2009entangled}%
  \BibitemOpen
  \bibfield  {author} {\bibinfo {author} {\bibfnamefont {J.~D.}\ \bibnamefont
  {Jost}}, \bibinfo {author} {\bibfnamefont {J.~P.}\ \bibnamefont {Home}},
  \bibinfo {author} {\bibfnamefont {J.~M.}\ \bibnamefont {Amini}}, \bibinfo
  {author} {\bibfnamefont {D.}~\bibnamefont {Hanneke}}, \bibinfo {author}
  {\bibfnamefont {R.}~\bibnamefont {Ozeri}}, \bibinfo {author} {\bibfnamefont
  {C.}~\bibnamefont {Langer}}, \bibinfo {author} {\bibfnamefont {J.~J.}\
  \bibnamefont {Bollinger}}, \bibinfo {author} {\bibfnamefont {D.}~\bibnamefont
  {Leibfried}}, \ and\ \bibinfo {author} {\bibfnamefont {D.~J.}\ \bibnamefont
  {Wineland}},\ }\href@noop {} {\bibfield  {journal} {\bibinfo  {journal}
  {Nature}\ }\textbf {\bibinfo {volume} {459}},\ \bibinfo {pages} {683}
  (\bibinfo {year} {2009})}\BibitemShut {NoStop}%
\bibitem [{\citenamefont {O’Connell}\ \emph {et~al.}(2010)\citenamefont
  {O’Connell}, \citenamefont {Hofheinz}, \citenamefont {Ansmann},
  \citenamefont {Bialczak}, \citenamefont {Lenander}, \citenamefont {Lucero},
  \citenamefont {Neeley}, \citenamefont {Sank}, \citenamefont {Wang},
  \citenamefont {Weides} \emph {et~al.}}]{o2010quantum}%
  \BibitemOpen
  \bibfield  {author} {\bibinfo {author} {\bibfnamefont {A.~D.}\ \bibnamefont
  {O’Connell}}, \bibinfo {author} {\bibfnamefont {M.}~\bibnamefont
  {Hofheinz}}, \bibinfo {author} {\bibfnamefont {M.}~\bibnamefont {Ansmann}},
  \bibinfo {author} {\bibfnamefont {R.~C.}\ \bibnamefont {Bialczak}}, \bibinfo
  {author} {\bibfnamefont {M.}~\bibnamefont {Lenander}}, \bibinfo {author}
  {\bibfnamefont {E.}~\bibnamefont {Lucero}}, \bibinfo {author} {\bibfnamefont
  {M.}~\bibnamefont {Neeley}}, \bibinfo {author} {\bibfnamefont
  {D.}~\bibnamefont {Sank}}, \bibinfo {author} {\bibfnamefont {H.}~\bibnamefont
  {Wang}}, \bibinfo {author} {\bibfnamefont {M.}~\bibnamefont {Weides}},  \emph
  {et~al.},\ }\href@noop {} {\bibfield  {journal} {\bibinfo  {journal}
  {Nature}\ }\textbf {\bibinfo {volume} {464}},\ \bibinfo {pages} {697}
  (\bibinfo {year} {2010})}\BibitemShut {NoStop}%
\bibitem [{\citenamefont {Aspelmeyer}\ \emph {et~al.}(2014)\citenamefont
  {Aspelmeyer}, \citenamefont {Kippenberg},\ and\ \citenamefont
  {Marquardt}}]{aspelmeyer2014cavity}%
  \BibitemOpen
  \bibfield  {author} {\bibinfo {author} {\bibfnamefont {M.}~\bibnamefont
  {Aspelmeyer}}, \bibinfo {author} {\bibfnamefont {T.~J.}\ \bibnamefont
  {Kippenberg}}, \ and\ \bibinfo {author} {\bibfnamefont {F.}~\bibnamefont
  {Marquardt}},\ }\href@noop {} {\bibfield  {journal} {\bibinfo  {journal}
  {Reviews of Modern Physics}\ }\textbf {\bibinfo {volume} {86}},\ \bibinfo
  {pages} {1391} (\bibinfo {year} {2014})}\BibitemShut {NoStop}%
\bibitem [{\citenamefont {LaHaye}\ \emph {et~al.}(2004)\citenamefont {LaHaye},
  \citenamefont {Buu}, \citenamefont {Camarota},\ and\ \citenamefont
  {Schwab}}]{lahaye2004approaching}%
  \BibitemOpen
  \bibfield  {author} {\bibinfo {author} {\bibfnamefont {M.}~\bibnamefont
  {LaHaye}}, \bibinfo {author} {\bibfnamefont {O.}~\bibnamefont {Buu}},
  \bibinfo {author} {\bibfnamefont {B.}~\bibnamefont {Camarota}}, \ and\
  \bibinfo {author} {\bibfnamefont {K.}~\bibnamefont {Schwab}},\ }\href@noop {}
  {\bibfield  {journal} {\bibinfo  {journal} {Science}\ }\textbf {\bibinfo
  {volume} {304}},\ \bibinfo {pages} {74} (\bibinfo {year} {2004})}\BibitemShut
  {NoStop}%
\bibitem [{\citenamefont {Cleland}\ and\ \citenamefont
  {Geller}(2004)}]{cleland2004superconducting}%
  \BibitemOpen
  \bibfield  {author} {\bibinfo {author} {\bibfnamefont {A.~N.}\ \bibnamefont
  {Cleland}}\ and\ \bibinfo {author} {\bibfnamefont {M.~R.}\ \bibnamefont
  {Geller}},\ }\href@noop {} {\bibfield  {journal} {\bibinfo  {journal}
  {Physical review letters}\ }\textbf {\bibinfo {volume} {93}},\ \bibinfo
  {pages} {070501} (\bibinfo {year} {2004})}\BibitemShut {NoStop}%
\bibitem [{\citenamefont {Wilson-Rae}\ \emph {et~al.}(2004)\citenamefont
  {Wilson-Rae}, \citenamefont {Zoller},\ and\ \citenamefont
  {Imamoḡlu}}]{wilson2004laser}%
  \BibitemOpen
  \bibfield  {author} {\bibinfo {author} {\bibfnamefont {I.}~\bibnamefont
  {Wilson-Rae}}, \bibinfo {author} {\bibfnamefont {P.}~\bibnamefont {Zoller}},
  \ and\ \bibinfo {author} {\bibfnamefont {A.}~\bibnamefont {Imamoḡlu}},\
  }\href@noop {} {\bibfield  {journal} {\bibinfo  {journal} {Physical Review
  Letters}\ }\textbf {\bibinfo {volume} {92}},\ \bibinfo {pages} {075507}
  (\bibinfo {year} {2004})}\BibitemShut {NoStop}%
\bibitem [{\citenamefont {Martin}\ \emph {et~al.}(2004)\citenamefont {Martin},
  \citenamefont {Shnirman}, \citenamefont {Tian},\ and\ \citenamefont
  {Zoller}}]{martin2004ground}%
  \BibitemOpen
  \bibfield  {author} {\bibinfo {author} {\bibfnamefont {I.}~\bibnamefont
  {Martin}}, \bibinfo {author} {\bibfnamefont {A.}~\bibnamefont {Shnirman}},
  \bibinfo {author} {\bibfnamefont {L.}~\bibnamefont {Tian}}, \ and\ \bibinfo
  {author} {\bibfnamefont {P.}~\bibnamefont {Zoller}},\ }\href@noop {}
  {\bibfield  {journal} {\bibinfo  {journal} {Physical Review B}\ }\textbf
  {\bibinfo {volume} {69}},\ \bibinfo {pages} {125339} (\bibinfo {year}
  {2004})}\BibitemShut {NoStop}%
\bibitem [{\citenamefont {Wilson-Rae}\ \emph {et~al.}(2007)\citenamefont
  {Wilson-Rae}, \citenamefont {Nooshi}, \citenamefont {Zwerger},\ and\
  \citenamefont {Kippenberg}}]{PhysRevLett.99.093901}%
  \BibitemOpen
  \bibfield  {author} {\bibinfo {author} {\bibfnamefont {I.}~\bibnamefont
  {Wilson-Rae}}, \bibinfo {author} {\bibfnamefont {N.}~\bibnamefont {Nooshi}},
  \bibinfo {author} {\bibfnamefont {W.}~\bibnamefont {Zwerger}}, \ and\
  \bibinfo {author} {\bibfnamefont {T.~J.}\ \bibnamefont {Kippenberg}},\ }\href
  {\doibase 10.1103/PhysRevLett.99.093901} {\bibfield  {journal} {\bibinfo
  {journal} {Phys. Rev. Lett.}\ }\textbf {\bibinfo {volume} {99}},\ \bibinfo
  {pages} {093901} (\bibinfo {year} {2007})}\BibitemShut {NoStop}%
\bibitem [{\citenamefont {Marquardt}\ \emph {et~al.}(2007)\citenamefont
  {Marquardt}, \citenamefont {Chen}, \citenamefont {Clerk},\ and\ \citenamefont
  {Girvin}}]{PhysRevLett.99.093902}%
  \BibitemOpen
  \bibfield  {author} {\bibinfo {author} {\bibfnamefont {F.}~\bibnamefont
  {Marquardt}}, \bibinfo {author} {\bibfnamefont {J.~P.}\ \bibnamefont {Chen}},
  \bibinfo {author} {\bibfnamefont {A.~A.}\ \bibnamefont {Clerk}}, \ and\
  \bibinfo {author} {\bibfnamefont {S.~M.}\ \bibnamefont {Girvin}},\ }\href
  {\doibase 10.1103/PhysRevLett.99.093902} {\bibfield  {journal} {\bibinfo
  {journal} {Phys. Rev. Lett.}\ }\textbf {\bibinfo {volume} {99}},\ \bibinfo
  {pages} {093902} (\bibinfo {year} {2007})}\BibitemShut {NoStop}%
\bibitem [{\citenamefont {Chan}\ \emph {et~al.}(2011)\citenamefont {Chan},
  \citenamefont {Alegre}, \citenamefont {Safavi-Naeini}, \citenamefont {Hill},
  \citenamefont {Krause}, \citenamefont {Gr{\"o}blacher}, \citenamefont
  {Aspelmeyer},\ and\ \citenamefont {Painter}}]{chan2011laser}%
  \BibitemOpen
  \bibfield  {author} {\bibinfo {author} {\bibfnamefont {J.}~\bibnamefont
  {Chan}}, \bibinfo {author} {\bibfnamefont {T.~M.}\ \bibnamefont {Alegre}},
  \bibinfo {author} {\bibfnamefont {A.~H.}\ \bibnamefont {Safavi-Naeini}},
  \bibinfo {author} {\bibfnamefont {J.~T.}\ \bibnamefont {Hill}}, \bibinfo
  {author} {\bibfnamefont {A.}~\bibnamefont {Krause}}, \bibinfo {author}
  {\bibfnamefont {S.}~\bibnamefont {Gr{\"o}blacher}}, \bibinfo {author}
  {\bibfnamefont {M.}~\bibnamefont {Aspelmeyer}}, \ and\ \bibinfo {author}
  {\bibfnamefont {O.}~\bibnamefont {Painter}},\ }\href@noop {} {\bibfield
  {journal} {\bibinfo  {journal} {Nature}\ }\textbf {\bibinfo {volume} {478}},\
  \bibinfo {pages} {89} (\bibinfo {year} {2011})}\BibitemShut {NoStop}%
\bibitem [{\citenamefont {Cirac}\ \emph {et~al.}(1992)\citenamefont {Cirac},
  \citenamefont {Blatt}, \citenamefont {Zoller},\ and\ \citenamefont
  {Phillips}}]{cirac1992laser}%
  \BibitemOpen
  \bibfield  {author} {\bibinfo {author} {\bibfnamefont {J.~I.}\ \bibnamefont
  {Cirac}}, \bibinfo {author} {\bibfnamefont {R.}~\bibnamefont {Blatt}},
  \bibinfo {author} {\bibfnamefont {P.}~\bibnamefont {Zoller}}, \ and\ \bibinfo
  {author} {\bibfnamefont {W.}~\bibnamefont {Phillips}},\ }\href@noop {}
  {\bibfield  {journal} {\bibinfo  {journal} {Physical Review A}\ }\textbf
  {\bibinfo {volume} {46}},\ \bibinfo {pages} {2668} (\bibinfo {year}
  {1992})}\BibitemShut {NoStop}%
\bibitem [{\citenamefont {Diedrich}\ \emph {et~al.}(1989)\citenamefont
  {Diedrich}, \citenamefont {Bergquist}, \citenamefont {Itano},\ and\
  \citenamefont {Wineland}}]{diedrich1989laser}%
  \BibitemOpen
  \bibfield  {author} {\bibinfo {author} {\bibfnamefont {F.}~\bibnamefont
  {Diedrich}}, \bibinfo {author} {\bibfnamefont {J.}~\bibnamefont {Bergquist}},
  \bibinfo {author} {\bibfnamefont {W.~M.}\ \bibnamefont {Itano}}, \ and\
  \bibinfo {author} {\bibfnamefont {D.}~\bibnamefont {Wineland}},\ }\href@noop
  {} {\bibfield  {journal} {\bibinfo  {journal} {Physical Review Letters}\
  }\textbf {\bibinfo {volume} {62}},\ \bibinfo {pages} {403} (\bibinfo {year}
  {1989})}\BibitemShut {NoStop}%
\bibitem [{\citenamefont {Kepesidis}\ \emph {et~al.}(2013)\citenamefont
  {Kepesidis}, \citenamefont {Bennett}, \citenamefont {Portolan}, \citenamefont
  {Lukin},\ and\ \citenamefont {Rabl}}]{kepesidis2013phonon}%
  \BibitemOpen
  \bibfield  {author} {\bibinfo {author} {\bibfnamefont {K.}~\bibnamefont
  {Kepesidis}}, \bibinfo {author} {\bibfnamefont {S.}~\bibnamefont {Bennett}},
  \bibinfo {author} {\bibfnamefont {S.}~\bibnamefont {Portolan}}, \bibinfo
  {author} {\bibfnamefont {M.~D.}\ \bibnamefont {Lukin}}, \ and\ \bibinfo
  {author} {\bibfnamefont {P.}~\bibnamefont {Rabl}},\ }\href@noop {} {\bibfield
   {journal} {\bibinfo  {journal} {Physical Review B}\ }\textbf {\bibinfo
  {volume} {88}},\ \bibinfo {pages} {064105} (\bibinfo {year}
  {2013})}\BibitemShut {NoStop}%
\bibitem [{\citenamefont {MacQuarrie}\ \emph {et~al.}(2013)\citenamefont
  {MacQuarrie}, \citenamefont {Gosavi}, \citenamefont {Jungwirth},
  \citenamefont {Bhave},\ and\ \citenamefont
  {Fuchs}}]{macquarrie2013mechanical}%
  \BibitemOpen
  \bibfield  {author} {\bibinfo {author} {\bibfnamefont {E.}~\bibnamefont
  {MacQuarrie}}, \bibinfo {author} {\bibfnamefont {T.}~\bibnamefont {Gosavi}},
  \bibinfo {author} {\bibfnamefont {N.}~\bibnamefont {Jungwirth}}, \bibinfo
  {author} {\bibfnamefont {S.}~\bibnamefont {Bhave}}, \ and\ \bibinfo {author}
  {\bibfnamefont {G.}~\bibnamefont {Fuchs}},\ }\href@noop {} {\bibfield
  {journal} {\bibinfo  {journal} {Physical review letters}\ }\textbf {\bibinfo
  {volume} {111}},\ \bibinfo {pages} {227602} (\bibinfo {year}
  {2013})}\BibitemShut {NoStop}%
\bibitem [{\citenamefont {MacQuarrie}\ \emph {et~al.}(2017)\citenamefont
  {MacQuarrie}, \citenamefont {Otten}, \citenamefont {Gray},\ and\
  \citenamefont {Fuchs}}]{macquarrie2017cooling}%
  \BibitemOpen
  \bibfield  {author} {\bibinfo {author} {\bibfnamefont {E.}~\bibnamefont
  {MacQuarrie}}, \bibinfo {author} {\bibfnamefont {M.}~\bibnamefont {Otten}},
  \bibinfo {author} {\bibfnamefont {S.}~\bibnamefont {Gray}}, \ and\ \bibinfo
  {author} {\bibfnamefont {G.}~\bibnamefont {Fuchs}},\ }\href@noop {}
  {\bibfield  {journal} {\bibinfo  {journal} {Nature Communications}\ }\textbf
  {\bibinfo {volume} {8}},\ \bibinfo {pages} {14358} (\bibinfo {year}
  {2017})}\BibitemShut {NoStop}%
\bibitem [{\citenamefont {Meekhof}\ \emph {et~al.}(1996)\citenamefont
  {Meekhof}, \citenamefont {Monroe}, \citenamefont {King}, \citenamefont
  {Itano},\ and\ \citenamefont {Wineland}}]{meekhof1996generation}%
  \BibitemOpen
  \bibfield  {author} {\bibinfo {author} {\bibfnamefont {D.}~\bibnamefont
  {Meekhof}}, \bibinfo {author} {\bibfnamefont {C.}~\bibnamefont {Monroe}},
  \bibinfo {author} {\bibfnamefont {B.}~\bibnamefont {King}}, \bibinfo {author}
  {\bibfnamefont {W.~M.}\ \bibnamefont {Itano}}, \ and\ \bibinfo {author}
  {\bibfnamefont {D.~J.}\ \bibnamefont {Wineland}},\ }\href@noop {} {\bibfield
  {journal} {\bibinfo  {journal} {Physical Review Letters}\ }\textbf {\bibinfo
  {volume} {76}},\ \bibinfo {pages} {1796} (\bibinfo {year}
  {1996})}\BibitemShut {NoStop}%
\bibitem [{\citenamefont {Cirac}\ and\ \citenamefont
  {Zoller}(1995)}]{cirac1995quantum}%
  \BibitemOpen
  \bibfield  {author} {\bibinfo {author} {\bibfnamefont {J.~I.}\ \bibnamefont
  {Cirac}}\ and\ \bibinfo {author} {\bibfnamefont {P.}~\bibnamefont {Zoller}},\
  }\href@noop {} {\bibfield  {journal} {\bibinfo  {journal} {Physical review
  letters}\ }\textbf {\bibinfo {volume} {74}},\ \bibinfo {pages} {4091}
  (\bibinfo {year} {1995})}\BibitemShut {NoStop}%
\bibitem [{\citenamefont {Monroe}\ \emph {et~al.}(1995)\citenamefont {Monroe},
  \citenamefont {Meekhof}, \citenamefont {King}, \citenamefont {Itano},\ and\
  \citenamefont {Wineland}}]{monroe1995demonstration}%
  \BibitemOpen
  \bibfield  {author} {\bibinfo {author} {\bibfnamefont {C.}~\bibnamefont
  {Monroe}}, \bibinfo {author} {\bibfnamefont {D.}~\bibnamefont {Meekhof}},
  \bibinfo {author} {\bibfnamefont {B.}~\bibnamefont {King}}, \bibinfo {author}
  {\bibfnamefont {W.~M.}\ \bibnamefont {Itano}}, \ and\ \bibinfo {author}
  {\bibfnamefont {D.~J.}\ \bibnamefont {Wineland}},\ }\href@noop {} {\bibfield
  {journal} {\bibinfo  {journal} {Physical review letters}\ }\textbf {\bibinfo
  {volume} {75}},\ \bibinfo {pages} {4714} (\bibinfo {year}
  {1995})}\BibitemShut {NoStop}%
\bibitem [{\citenamefont {Armour}\ \emph {et~al.}(2002)\citenamefont {Armour},
  \citenamefont {Blencowe},\ and\ \citenamefont
  {Schwab}}]{armour2002entanglement}%
  \BibitemOpen
  \bibfield  {author} {\bibinfo {author} {\bibfnamefont {A.}~\bibnamefont
  {Armour}}, \bibinfo {author} {\bibfnamefont {M.}~\bibnamefont {Blencowe}}, \
  and\ \bibinfo {author} {\bibfnamefont {K.~C.}\ \bibnamefont {Schwab}},\
  }\href@noop {} {\bibfield  {journal} {\bibinfo  {journal} {Physical Review
  Letters}\ }\textbf {\bibinfo {volume} {88}},\ \bibinfo {pages} {148301}
  (\bibinfo {year} {2002})}\BibitemShut {NoStop}%
\bibitem [{\citenamefont {Cirac}\ \emph {et~al.}(1991)\citenamefont {Cirac},
  \citenamefont {Ritsch},\ and\ \citenamefont {Zoller}}]{cirac1991two}%
  \BibitemOpen
  \bibfield  {author} {\bibinfo {author} {\bibfnamefont {J.}~\bibnamefont
  {Cirac}}, \bibinfo {author} {\bibfnamefont {H.}~\bibnamefont {Ritsch}}, \
  and\ \bibinfo {author} {\bibfnamefont {P.}~\bibnamefont {Zoller}},\
  }\href@noop {} {\bibfield  {journal} {\bibinfo  {journal} {Physical Review
  A}\ }\textbf {\bibinfo {volume} {44}},\ \bibinfo {pages} {4541} (\bibinfo
  {year} {1991})}\BibitemShut {NoStop}%
\bibitem [{\citenamefont {Yariv}(2000)}]{yariv2000universal}%
  \BibitemOpen
  \bibfield  {author} {\bibinfo {author} {\bibfnamefont {A.}~\bibnamefont
  {Yariv}},\ }\href@noop {} {\bibfield  {journal} {\bibinfo  {journal}
  {Electronics letters}\ }\textbf {\bibinfo {volume} {36}},\ \bibinfo {pages}
  {321} (\bibinfo {year} {2000})}\BibitemShut {NoStop}%
\bibitem [{\citenamefont {Piper}\ and\ \citenamefont
  {Fan}(2014)}]{piper2014total}%
  \BibitemOpen
  \bibfield  {author} {\bibinfo {author} {\bibfnamefont {J.~R.}\ \bibnamefont
  {Piper}}\ and\ \bibinfo {author} {\bibfnamefont {S.}~\bibnamefont {Fan}},\
  }\href@noop {} {\bibfield  {journal} {\bibinfo  {journal} {Acs Photonics}\
  }\textbf {\bibinfo {volume} {1}},\ \bibinfo {pages} {347} (\bibinfo {year}
  {2014})}\BibitemShut {NoStop}%
\bibitem [{\citenamefont {Jeon}\ \emph {et~al.}()\citenamefont {Jeon},
  \citenamefont {Hernandez}, \citenamefont {Rosenmann}, \citenamefont {Gray},
  \citenamefont {Martinson},\ and\ \citenamefont {Foley~IV}}]{jeon2018pareto}%
  \BibitemOpen
  \bibfield  {author} {\bibinfo {author} {\bibfnamefont {N.}~\bibnamefont
  {Jeon}}, \bibinfo {author} {\bibfnamefont {J.~J.}\ \bibnamefont {Hernandez}},
  \bibinfo {author} {\bibfnamefont {D.}~\bibnamefont {Rosenmann}}, \bibinfo
  {author} {\bibfnamefont {S.~K.}\ \bibnamefont {Gray}}, \bibinfo {author}
  {\bibfnamefont {A.~B.}\ \bibnamefont {Martinson}}, \ and\ \bibinfo {author}
  {\bibfnamefont {J.~J.}\ \bibnamefont {Foley~IV}},\ }\href@noop {} {\bibinfo
  {journal} {Advanced Energy Materials}\ ,\ \bibinfo {pages}
  {1801035}}\BibitemShut {NoStop}%
\bibitem [{\citenamefont {Zanotto}\ \emph {et~al.}(2014)\citenamefont
  {Zanotto}, \citenamefont {Mezzapesa}, \citenamefont {Bianco}, \citenamefont
  {Biasiol}, \citenamefont {Baldacci}, \citenamefont {Vitiello}, \citenamefont
  {Sorba}, \citenamefont {Colombelli},\ and\ \citenamefont
  {Tredicucci}}]{zanotto2014perfect}%
  \BibitemOpen
\bibfield  {journal} {  }\bibfield  {author} {\bibinfo {author} {\bibfnamefont
  {S.}~\bibnamefont {Zanotto}}, \bibinfo {author} {\bibfnamefont {F.~P.}\
  \bibnamefont {Mezzapesa}}, \bibinfo {author} {\bibfnamefont {F.}~\bibnamefont
  {Bianco}}, \bibinfo {author} {\bibfnamefont {G.}~\bibnamefont {Biasiol}},
  \bibinfo {author} {\bibfnamefont {L.}~\bibnamefont {Baldacci}}, \bibinfo
  {author} {\bibfnamefont {M.~S.}\ \bibnamefont {Vitiello}}, \bibinfo {author}
  {\bibfnamefont {L.}~\bibnamefont {Sorba}}, \bibinfo {author} {\bibfnamefont
  {R.}~\bibnamefont {Colombelli}}, \ and\ \bibinfo {author} {\bibfnamefont
  {A.}~\bibnamefont {Tredicucci}},\ }\href@noop {} {\bibfield  {journal}
  {\bibinfo  {journal} {Nature Physics}\ }\textbf {\bibinfo {volume} {10}},\
  \bibinfo {pages} {830} (\bibinfo {year} {2014})}\BibitemShut {NoStop}%
\bibitem [{\citenamefont {Cai}\ \emph {et~al.}(2000)\citenamefont {Cai},
  \citenamefont {Painter},\ and\ \citenamefont {Vahala}}]{cai2000observation}%
  \BibitemOpen
  \bibfield  {author} {\bibinfo {author} {\bibfnamefont {M.}~\bibnamefont
  {Cai}}, \bibinfo {author} {\bibfnamefont {O.}~\bibnamefont {Painter}}, \ and\
  \bibinfo {author} {\bibfnamefont {K.~J.}\ \bibnamefont {Vahala}},\
  }\href@noop {} {\bibfield  {journal} {\bibinfo  {journal} {Physical review
  letters}\ }\textbf {\bibinfo {volume} {85}},\ \bibinfo {pages} {74} (\bibinfo
  {year} {2000})}\BibitemShut {NoStop}%
\bibitem [{\citenamefont {Lee}\ \emph {et~al.}(2017)\citenamefont {Lee},
  \citenamefont {Lee}, \citenamefont {Cady}, \citenamefont {Ovartchaiyapong},\
  and\ \citenamefont {Jayich}}]{lee2017topical}%
  \BibitemOpen
  \bibfield  {author} {\bibinfo {author} {\bibfnamefont {D.}~\bibnamefont
  {Lee}}, \bibinfo {author} {\bibfnamefont {K.~W.}\ \bibnamefont {Lee}},
  \bibinfo {author} {\bibfnamefont {J.~V.}\ \bibnamefont {Cady}}, \bibinfo
  {author} {\bibfnamefont {P.}~\bibnamefont {Ovartchaiyapong}}, \ and\ \bibinfo
  {author} {\bibfnamefont {A.~C.~B.}\ \bibnamefont {Jayich}},\ }\href@noop {}
  {\bibfield  {journal} {\bibinfo  {journal} {Journal of Optics}\ }\textbf
  {\bibinfo {volume} {19}},\ \bibinfo {pages} {033001} (\bibinfo {year}
  {2017})}\BibitemShut {NoStop}%
\bibitem [{\citenamefont {Carmele}\ \emph {et~al.}(2014)\citenamefont
  {Carmele}, \citenamefont {Vogell}, \citenamefont {Stannigel},\ and\
  \citenamefont {Zoller}}]{carmele2014opto}%
  \BibitemOpen
  \bibfield  {author} {\bibinfo {author} {\bibfnamefont {A.}~\bibnamefont
  {Carmele}}, \bibinfo {author} {\bibfnamefont {B.}~\bibnamefont {Vogell}},
  \bibinfo {author} {\bibfnamefont {K.}~\bibnamefont {Stannigel}}, \ and\
  \bibinfo {author} {\bibfnamefont {P.}~\bibnamefont {Zoller}},\ }\href@noop {}
  {\bibfield  {journal} {\bibinfo  {journal} {New Journal of Physics}\ }\textbf
  {\bibinfo {volume} {16}},\ \bibinfo {pages} {063042} (\bibinfo {year}
  {2014})}\BibitemShut {NoStop}%
\bibitem [{\citenamefont {Ficek}\ and\ \citenamefont
  {Tana{\'s}}(2002)}]{ficek2002entangled}%
  \BibitemOpen
  \bibfield  {author} {\bibinfo {author} {\bibfnamefont {Z.}~\bibnamefont
  {Ficek}}\ and\ \bibinfo {author} {\bibfnamefont {R.}~\bibnamefont
  {Tana{\'s}}},\ }\href@noop {} {\bibfield  {journal} {\bibinfo  {journal}
  {Physics Reports}\ }\textbf {\bibinfo {volume} {372}},\ \bibinfo {pages}
  {369} (\bibinfo {year} {2002})}\BibitemShut {NoStop}%
\bibitem [{\citenamefont {Dung}\ \emph {et~al.}(2002)\citenamefont {Dung},
  \citenamefont {Kn{\"o}ll},\ and\ \citenamefont {Welsch}}]{dung2002resonant}%
  \BibitemOpen
  \bibfield  {author} {\bibinfo {author} {\bibfnamefont {H.~T.}\ \bibnamefont
  {Dung}}, \bibinfo {author} {\bibfnamefont {L.}~\bibnamefont {Kn{\"o}ll}}, \
  and\ \bibinfo {author} {\bibfnamefont {D.-G.}\ \bibnamefont {Welsch}},\
  }\href@noop {} {\bibfield  {journal} {\bibinfo  {journal} {Physical Review
  A}\ }\textbf {\bibinfo {volume} {66}},\ \bibinfo {pages} {063810} (\bibinfo
  {year} {2002})}\BibitemShut {NoStop}%
\bibitem [{\citenamefont {Varada}\ and\ \citenamefont
  {Agarwal}(1992)}]{varada1992two}%
  \BibitemOpen
  \bibfield  {author} {\bibinfo {author} {\bibfnamefont {G.}~\bibnamefont
  {Varada}}\ and\ \bibinfo {author} {\bibfnamefont {G.}~\bibnamefont
  {Agarwal}},\ }\href@noop {} {\bibfield  {journal} {\bibinfo  {journal}
  {Physical Review A}\ }\textbf {\bibinfo {volume} {45}},\ \bibinfo {pages}
  {6721} (\bibinfo {year} {1992})}\BibitemShut {NoStop}%
\bibitem [{\citenamefont {Van~Loo}\ \emph {et~al.}(2013)\citenamefont
  {Van~Loo}, \citenamefont {Fedorov}, \citenamefont {Lalumi{\`e}re},
  \citenamefont {Sanders}, \citenamefont {Blais},\ and\ \citenamefont
  {Wallraff}}]{van2013photon}%
  \BibitemOpen
  \bibfield  {author} {\bibinfo {author} {\bibfnamefont {A.~F.}\ \bibnamefont
  {Van~Loo}}, \bibinfo {author} {\bibfnamefont {A.}~\bibnamefont {Fedorov}},
  \bibinfo {author} {\bibfnamefont {K.}~\bibnamefont {Lalumi{\`e}re}}, \bibinfo
  {author} {\bibfnamefont {B.~C.}\ \bibnamefont {Sanders}}, \bibinfo {author}
  {\bibfnamefont {A.}~\bibnamefont {Blais}}, \ and\ \bibinfo {author}
  {\bibfnamefont {A.}~\bibnamefont {Wallraff}},\ }\href@noop {} {\bibfield
  {journal} {\bibinfo  {journal} {Science}\ }\textbf {\bibinfo {volume}
  {342}},\ \bibinfo {pages} {1494} (\bibinfo {year} {2013})}\BibitemShut
  {NoStop}%
\bibitem [{\citenamefont {Mlynek}\ \emph {et~al.}(2014)\citenamefont {Mlynek},
  \citenamefont {Abdumalikov}, \citenamefont {Eichler},\ and\ \citenamefont
  {Wallraff}}]{mlynek2014observation}%
  \BibitemOpen
  \bibfield  {author} {\bibinfo {author} {\bibfnamefont {J.}~\bibnamefont
  {Mlynek}}, \bibinfo {author} {\bibfnamefont {A.}~\bibnamefont {Abdumalikov}},
  \bibinfo {author} {\bibfnamefont {C.}~\bibnamefont {Eichler}}, \ and\
  \bibinfo {author} {\bibfnamefont {A.}~\bibnamefont {Wallraff}},\ }\href@noop
  {} {\bibfield  {journal} {\bibinfo  {journal} {Nature Communications}\
  }\textbf {\bibinfo {volume} {5}},\ \bibinfo {pages} {5186} (\bibinfo {year}
  {2014})}\BibitemShut {NoStop}%
\bibitem [{\citenamefont {Metelmann}\ and\ \citenamefont
  {Clerk}(2015)}]{metelmann2015nonreciprocal}%
  \BibitemOpen
  \bibfield  {author} {\bibinfo {author} {\bibfnamefont {A.}~\bibnamefont
  {Metelmann}}\ and\ \bibinfo {author} {\bibfnamefont {A.~A.}\ \bibnamefont
  {Clerk}},\ }\href@noop {} {\bibfield  {journal} {\bibinfo  {journal}
  {Physical Review X}\ }\textbf {\bibinfo {volume} {5}},\ \bibinfo {pages}
  {021025} (\bibinfo {year} {2015})}\BibitemShut {NoStop}%
\bibitem [{\citenamefont {Cortes}\ and\ \citenamefont
  {Jacob}(2017)}]{cortes2017super}%
  \BibitemOpen
  \bibfield  {author} {\bibinfo {author} {\bibfnamefont {C.~L.}\ \bibnamefont
  {Cortes}}\ and\ \bibinfo {author} {\bibfnamefont {Z.}~\bibnamefont {Jacob}},\
  }\href@noop {} {\bibfield  {journal} {\bibinfo  {journal} {Nature
  Communications}\ }\textbf {\bibinfo {volume} {8}},\ \bibinfo {pages} {14144}
  (\bibinfo {year} {2017})}\BibitemShut {NoStop}%
\bibitem [{\citenamefont {Newman}\ \emph {et~al.}(2018)\citenamefont {Newman},
  \citenamefont {Cortes}, \citenamefont {Afshar}, \citenamefont {Cadien},
  \citenamefont {Meldrum}, \citenamefont {Fedosejevs},\ and\ \citenamefont
  {Jacob}}]{newman2018observation}%
  \BibitemOpen
  \bibfield  {author} {\bibinfo {author} {\bibfnamefont {W.~D.}\ \bibnamefont
  {Newman}}, \bibinfo {author} {\bibfnamefont {C.~L.}\ \bibnamefont {Cortes}},
  \bibinfo {author} {\bibfnamefont {A.}~\bibnamefont {Afshar}}, \bibinfo
  {author} {\bibfnamefont {K.}~\bibnamefont {Cadien}}, \bibinfo {author}
  {\bibfnamefont {A.}~\bibnamefont {Meldrum}}, \bibinfo {author} {\bibfnamefont
  {R.}~\bibnamefont {Fedosejevs}}, \ and\ \bibinfo {author} {\bibfnamefont
  {Z.}~\bibnamefont {Jacob}},\ }\href@noop {} {\bibfield  {journal} {\bibinfo
  {journal} {Science Advances}\ }\textbf {\bibinfo {volume} {4}},\ \bibinfo
  {pages} {eaar5278} (\bibinfo {year} {2018})}\BibitemShut {NoStop}%
\bibitem [{\citenamefont {Mahmoud}\ \emph {et~al.}(2017)\citenamefont
  {Mahmoud}, \citenamefont {Liberal},\ and\ \citenamefont
  {Engheta}}]{mahmoud2017dipole}%
  \BibitemOpen
  \bibfield  {author} {\bibinfo {author} {\bibfnamefont {A.}~\bibnamefont
  {Mahmoud}}, \bibinfo {author} {\bibfnamefont {I.}~\bibnamefont {Liberal}}, \
  and\ \bibinfo {author} {\bibfnamefont {N.}~\bibnamefont {Engheta}},\
  }\href@noop {} {\bibfield  {journal} {\bibinfo  {journal} {Optical Materials
  Express}\ }\textbf {\bibinfo {volume} {7}},\ \bibinfo {pages} {415} (\bibinfo
  {year} {2017})}\BibitemShut {NoStop}%
\bibitem [{\citenamefont {Dicke}(1954)}]{dicke1954coherence}%
  \BibitemOpen
  \bibfield  {author} {\bibinfo {author} {\bibfnamefont {R.~H.}\ \bibnamefont
  {Dicke}},\ }\href@noop {} {\bibfield  {journal} {\bibinfo  {journal}
  {Physical Review}\ }\textbf {\bibinfo {volume} {93}},\ \bibinfo {pages} {99}
  (\bibinfo {year} {1954})}\BibitemShut {NoStop}%
\bibitem [{\citenamefont {Ovartchaiyapong}\ \emph {et~al.}(2014)\citenamefont
  {Ovartchaiyapong}, \citenamefont {Lee}, \citenamefont {Myers},\ and\
  \citenamefont {Jayich}}]{ovartchaiyapong2014dynamic}%
  \BibitemOpen
  \bibfield  {author} {\bibinfo {author} {\bibfnamefont {P.}~\bibnamefont
  {Ovartchaiyapong}}, \bibinfo {author} {\bibfnamefont {K.~W.}\ \bibnamefont
  {Lee}}, \bibinfo {author} {\bibfnamefont {B.~A.}\ \bibnamefont {Myers}}, \
  and\ \bibinfo {author} {\bibfnamefont {A.~C.~B.}\ \bibnamefont {Jayich}},\
  }\href@noop {} {\bibfield  {journal} {\bibinfo  {journal} {Nature
  Communications}\ }\textbf {\bibinfo {volume} {5}},\ \bibinfo {pages} {4429}
  (\bibinfo {year} {2014})}\BibitemShut {NoStop}%
\bibitem [{\citenamefont {Meesala}\ \emph {et~al.}(2016)\citenamefont
  {Meesala}, \citenamefont {Sohn}, \citenamefont {Atikian}, \citenamefont
  {Kim}, \citenamefont {Burek}, \citenamefont {Choy},\ and\ \citenamefont
  {Lon{\v{c}}ar}}]{meesala2016enhanced}%
  \BibitemOpen
  \bibfield  {author} {\bibinfo {author} {\bibfnamefont {S.}~\bibnamefont
  {Meesala}}, \bibinfo {author} {\bibfnamefont {Y.-I.}\ \bibnamefont {Sohn}},
  \bibinfo {author} {\bibfnamefont {H.~A.}\ \bibnamefont {Atikian}}, \bibinfo
  {author} {\bibfnamefont {S.}~\bibnamefont {Kim}}, \bibinfo {author}
  {\bibfnamefont {M.~J.}\ \bibnamefont {Burek}}, \bibinfo {author}
  {\bibfnamefont {J.~T.}\ \bibnamefont {Choy}}, \ and\ \bibinfo {author}
  {\bibfnamefont {M.}~\bibnamefont {Lon{\v{c}}ar}},\ }\href@noop {} {\bibfield
  {journal} {\bibinfo  {journal} {Physical Review Applied}\ }\textbf {\bibinfo
  {volume} {5}},\ \bibinfo {pages} {034010} (\bibinfo {year}
  {2016})}\BibitemShut {NoStop}%
\bibitem [{\citenamefont {Sohn}\ \emph {et~al.}(2018)\citenamefont {Sohn},
  \citenamefont {Meesala}, \citenamefont {Pingault}, \citenamefont {Atikian},
  \citenamefont {Holzgrafe}, \citenamefont {G{\"u}ndo{\u{g}}an}, \citenamefont
  {Stavrakas}, \citenamefont {Stanley}, \citenamefont {Sipahigil},
  \citenamefont {Choi} \emph {et~al.}}]{sohn2018controlling}%
  \BibitemOpen
  \bibfield  {author} {\bibinfo {author} {\bibfnamefont {Y.-I.}\ \bibnamefont
  {Sohn}}, \bibinfo {author} {\bibfnamefont {S.}~\bibnamefont {Meesala}},
  \bibinfo {author} {\bibfnamefont {B.}~\bibnamefont {Pingault}}, \bibinfo
  {author} {\bibfnamefont {H.~A.}\ \bibnamefont {Atikian}}, \bibinfo {author}
  {\bibfnamefont {J.}~\bibnamefont {Holzgrafe}}, \bibinfo {author}
  {\bibfnamefont {M.}~\bibnamefont {G{\"u}ndo{\u{g}}an}}, \bibinfo {author}
  {\bibfnamefont {C.}~\bibnamefont {Stavrakas}}, \bibinfo {author}
  {\bibfnamefont {M.~J.}\ \bibnamefont {Stanley}}, \bibinfo {author}
  {\bibfnamefont {A.}~\bibnamefont {Sipahigil}}, \bibinfo {author}
  {\bibfnamefont {J.}~\bibnamefont {Choi}},  \emph {et~al.},\ }\href@noop {}
  {\bibfield  {journal} {\bibinfo  {journal} {Nature Communications}\ }\textbf
  {\bibinfo {volume} {9}},\ \bibinfo {pages} {2012} (\bibinfo {year}
  {2018})}\BibitemShut {NoStop}%
\bibitem [{\citenamefont {Wolfe}\ and\ \citenamefont
  {Yelin}(2014)}]{wolfe2014certifying}%
  \BibitemOpen
  \bibfield  {author} {\bibinfo {author} {\bibfnamefont {E.}~\bibnamefont
  {Wolfe}}\ and\ \bibinfo {author} {\bibfnamefont {S.}~\bibnamefont {Yelin}},\
  }\href@noop {} {\bibfield  {journal} {\bibinfo  {journal} {Physical review
  letters}\ }\textbf {\bibinfo {volume} {112}},\ \bibinfo {pages} {140402}
  (\bibinfo {year} {2014})}\BibitemShut {NoStop}%
\bibitem [{\citenamefont {Otten}\ \emph {et~al.}(2016)\citenamefont {Otten},
  \citenamefont {Larson}, \citenamefont {Min}, \citenamefont {Wild},
  \citenamefont {Pelton},\ and\ \citenamefont {Gray}}]{otten2016origins}%
  \BibitemOpen
  \bibfield  {author} {\bibinfo {author} {\bibfnamefont {M.}~\bibnamefont
  {Otten}}, \bibinfo {author} {\bibfnamefont {J.}~\bibnamefont {Larson}},
  \bibinfo {author} {\bibfnamefont {M.}~\bibnamefont {Min}}, \bibinfo {author}
  {\bibfnamefont {S.~M.}\ \bibnamefont {Wild}}, \bibinfo {author}
  {\bibfnamefont {M.}~\bibnamefont {Pelton}}, \ and\ \bibinfo {author}
  {\bibfnamefont {S.~K.}\ \bibnamefont {Gray}},\ }\href@noop {} {\bibfield
  {journal} {\bibinfo  {journal} {Physical Review A}\ }\textbf {\bibinfo
  {volume} {94}},\ \bibinfo {pages} {022312} (\bibinfo {year}
  {2016})}\BibitemShut {NoStop}%
\bibitem [{\citenamefont {Otten}\ \emph {et~al.}(2015)\citenamefont {Otten},
  \citenamefont {Shah}, \citenamefont {Scherer}, \citenamefont {Min},
  \citenamefont {Pelton},\ and\ \citenamefont {Gray}}]{otten2015entanglement}%
  \BibitemOpen
  \bibfield  {author} {\bibinfo {author} {\bibfnamefont {M.}~\bibnamefont
  {Otten}}, \bibinfo {author} {\bibfnamefont {R.~A.}\ \bibnamefont {Shah}},
  \bibinfo {author} {\bibfnamefont {N.~F.}\ \bibnamefont {Scherer}}, \bibinfo
  {author} {\bibfnamefont {M.}~\bibnamefont {Min}}, \bibinfo {author}
  {\bibfnamefont {M.}~\bibnamefont {Pelton}}, \ and\ \bibinfo {author}
  {\bibfnamefont {S.~K.}\ \bibnamefont {Gray}},\ }\href@noop {} {\bibfield
  {journal} {\bibinfo  {journal} {Physical Review B}\ }\textbf {\bibinfo
  {volume} {92}},\ \bibinfo {pages} {125432} (\bibinfo {year}
  {2015})}\BibitemShut {NoStop}%
\bibitem [{\citenamefont {Temnov}\ and\ \citenamefont
  {Woggon}(2005)}]{temnov2005superradiance}%
  \BibitemOpen
  \bibfield  {author} {\bibinfo {author} {\bibfnamefont {V.~V.}\ \bibnamefont
  {Temnov}}\ and\ \bibinfo {author} {\bibfnamefont {U.}~\bibnamefont
  {Woggon}},\ }\href@noop {} {\bibfield  {journal} {\bibinfo  {journal}
  {Physical review letters}\ }\textbf {\bibinfo {volume} {95}},\ \bibinfo
  {pages} {243602} (\bibinfo {year} {2005})}\BibitemShut {NoStop}%
\end{thebibliography}


\end{document}